\documentclass{emulateapj}
\usepackage{bm}% bold math

\usepackage{fix-cm}
\usepackage{Rotating} % for rotating a large table.
\def\myfig#1{#1}

\newcounter{CommentNumber}
\newcommand{\TeXComment}[1]{{\textcolor{blue}{\tiny$^{[\theCommentNumber]}$}\stepcounter{CommentNumber}}}
\newcommand{\cm}{\mbox{ cm}}
\newcommand{\km}{\mbox{ km}}

\newcommand{\kpc}{\mbox{ kpc}}
\newcommand{\Mpc}{\mbox{ Mpc}}
\newcommand{\se}{\mbox{ s}}
\newcommand{\yr}{\mbox{ yr}}
\newcommand{\Gyr}{\mbox{ Gyr}}
\newcommand{\Hz}{\mbox{ Hz}}
\newcommand{\MHz}{\mbox{ MHz}}
\newcommand{\GHz}{\mbox{ GHz}}
\newcommand{\erg}{\mbox{ erg}}

\newcommand{\keV}{\mbox{ keV}}

\newcommand{\GeV}{\mbox{ GeV}}
\newcommand{\TeV}{\mbox{ TeV}}
\newcommand{\muG}{\mbox{ $\mu$G}}
\newcommand{\mb}{\mbox{ mb}}

\newcommand{\fin}{\mbox{ .}}
\newcommand{\coma}{\mbox{ ,}}
\newcommand{\ie}{\emph{i.e.} }
\newcommand{\eg}{\emph{e.g.,} }
\newcommand{\cf}{\emph{cf.} }

\newcommand{\ave}[1]{{\langle{#1}\rangle}}

\newcommand{\mypP}{{\kappa}}
\newcommand{\mypPP}{{\mypP_P}}
\newcommand{\mypT}{{\mypP}}

\newcommand{\mypB}{{\mypP_B}}

\newcommand{\myeta}{{\eta}}

\newcommand{\constant}{\mbox{constant}}

\newcommand{\CRP}{{CRP}}
\newcommand{\CRE}{{CRE}}
\newcommand{\CRI}{{CRI}}

\newcommand{\CRPs}{{CRPs}}
\newcommand{\CREs}{{CREs}}
\newcommand{\CRIs}{{CRIs}}

\newcommand{\AGNs}{{AGNs}}

\shorttitle{Modeling radio halos and minihalos}

\shortauthors{Keshet \& Loeb}

\begin{document}

\title{Using Radio Halos and Minihalos to Measure the Distributions \\
of Magnetic Fields and Cosmic-Rays in Galaxy Clusters}

\author{Uri Keshet\altaffilmark{1} and Abraham Loeb}

\affil{Harvard-Smithsonian Center for Astrophysics, 60 Garden St.,  Cambridge, MA 02138, USA}

\altaffiltext{1}{Einstein fellow}

\begin{abstract}
%arXiv abstract limit: 24 lines of 80 chars
%2345678901234567890123456789012345678901234567890123456789012345678901234567890
Some galaxy clusters show diffuse radio emission in the form of giant
halos (GHs) on Mpc scales or minihalos (MHs) on smaller
scales. Comparing Very Large Array and \emph{XMM-Newton} radial profiles of several such
clusters, we find a universal linear correlation between radio and
X-ray surface brightness, valid in both types of halos. It implies a halo central emissivity $\nu j_\nu = 10^{-31.4\pm0.2} (n/10^{-2}\cm^{-3})^2 (T/T_0)^{0.2\pm0.5} \erg\se^{-1}\cm^{-3}$, where $T$ and $T_0$ are the local and central temperatures, and $n$ is the electron number density.
We argue that the tight correlation and the scaling of $j_\nu$,
combined with morphological and spectral evidence, indicate that both
GHs and MHs arise from secondary electrons and positrons, produced in
cosmic-ray ion (CRI) collisions with a strongly magnetized, $B\gtrsim 3\muG$ intracluster gas.
When the magnetic energy density drops below that of the microwave
background, the radio emission weakens considerably, producing halos
with a clumpy morphology (\eg RXC\,J2003.5–-2323 and A2255) or a
distinct radial break.  We thus measure a magnetic field $B=3\muG$ at a radius
$r \simeq 110 \kpc$ in A2029 and $r\simeq 50\kpc$ in Perseus.  The
spectrum of secondaries, produced from hadronic collisions of
$\sim20\GeV$ {\CRIs}, reflects the energy dependence
of the collision cross section.
We use the observed spectra of halos, in particular where they steepen
with increasing radius or frequency, to \emph{(i)} measure $B\simeq
10(\nu/700\MHz)\muG$, with $\nu$ the spectral break frequency;
\emph{(ii)} identify a correlation between the average spectrum and
the central magnetic field; and \emph{(iii)} infer a {\CRI} spectral
index $s\lesssim -2.7$ and energy fraction $\xi_p\sim 10^{-3.6\pm0.2}$ at
particle energies above 10 GeV.  Our results favor a model where {\CRIs} diffuse
away from their sources (which are probably supernovae, according to a
preliminary correlation with star formation), whereas the
magnetic fields are generated by mergers in GHs and by core sloshing
in MHs.
\end{abstract}

\keywords{galaxies: clusters: general --- intergalactic medium --- X-rays: galaxies: clusters --- radio continuum: general --- magnetic fields}

\maketitle

% kticnyhfor jgntofefeo jfdeoamzit

\section{Introduction}

Giant halos (GHs) appear as diffuse radio emission on $\sim\mbox{Mpc}$
scales in merging galaxy clusters \cite[for a review,
see][]{FerettiGiovannini08}.  GHs were identified in about a quarter of all
clusters with X-ray luminosities $L_X>5\times 10^{44}\erg \se^{-1}$ at
redshifts $0.2<z<0.4$ \citep{BrunettiEtAl07}.  $L_X$ and the specific
radio power $P_\nu$ of GH clusters are tightly correlated. The GH
distribution in clusters is bimodal, with most clusters showing no
associated GH at a sensitivity threshold $\sim 10$ times better than the signal expected from the $L_X$--$P_\nu$ correlation \citep{BrunettiEtAl07}.

GHs arise from synchrotron radiation, emitted by cosmic-ray electrons and positrons ({\CREs}) injected locally and continuously into the magnetized plasma.
Such {\CREs} can lose a considerable fraction of their energy by inverse-Compton scattering off the cosmic microwave background (CMB).
Recently, \citet[][henceforth K09]{KushnirEtAl09} made
the important observation that the GH properties mentioned above are reproduced if the {\CREs}
are produced in $p$-$p$ cosmic ray proton ({\CRP})
collisions, and the magnetic field $B$ is sufficiently strong to saturate
$P_\nu(B)$ by rendering Compton losses negligible.

To qualitatively see this, assume that the {\CRP} number density $n_{p}$ is (narrowly distributed about) a universal fraction of the local (non cosmic-ray) electron number density $n$, and that the magnetic energy density greatly exceeds the energy density $u_{cmb}$ of the CMB, $B\gg B_{cmb}\equiv (8\pi u_{cmb})^{1/2}$.
Here, the ratio between the emissivities of radio synchrotron
$j_\nu \propto n_{p} n B^2/(B^2+B_{cmb}^2) \propto n^2$ (from secondary, density $n_e\propto n_p n$ {CREs})
and X-ray bremsstrahlung $j_X \propto n^2$ (from the thermal plasma),
does not depend on $n$ or on $B$.
Thus, strongly magnetized clusters with $B\gtrsim B_{cmb}$ would show a strong radio-X-ray correlation, whereas clusters with $B\lesssim B_{cmb}/3$ would be too radio faint to show a GH.
Other models, notably turbulent reacceleration of electrons \citep[for a review see][]{PetrosianBykov08}, do not naturally produce the correlation and
bimodality that are observed\footnote{In the sense that the small dispersion above $P_\nu(L_X)$ is not reproduced, and many assumptions \citep[\eg][]{BrunettiLazarian07} are needed.
For a different opinion, see \citet{BrunettiEtAl09}.}.

Radio minihalos (MHs) are found in cool core clusters (CCs).  They
extend roughly over the cooling region \citep{GittiEtAl02}, encompassing
up to a few percent of the typical GH volume.  Detecting MHs is more challenging than GHs, due to their smaller size and proximity to an active galactic nucleus (AGN), so only few MHs have been well studied and less is known about their correlation with X-rays.
A morphological association between MH edges and cold fronts
\citep[CFs;][]{MazzottaGiacintucci08} suggests a link between MHs and
sloshing activity in the core.  Such CFs are observed in about half the CCs, and are probably present in many more \citep{MarkevitchVikhlinin07}.  They
are thought to be tangential discontinuities that isolate regions
magnetized by bulk shear flow at smaller radii \citep{KeshetEtAl10}
(regions we refer to as below, or inside the CF).

There are many similarities between GHs and MHs.
Both types of halos are usually characterized by a regular morphology,
low surface brightness, little or no polarization,
and spectral indices $\alpha_\nu \equiv d\log(P_\nu)/d\log\nu\simeq
-(1.0-1.5)$, with $\nu$ the frequency.
However, there are telling exceptions to these characteristics.
A few GHs have a clumpy or filamentary morphology, such as in RXC\,J2003.5–-2323 \citep{GiacintucciEtAl09}, A2255, and A2319 \citep[][henceforth M09]{MurgiaEtAl09}.
Strong polarization was so far detected in one GH (at a $20\%$--$40\%$ level, in A2255; see \citet{GovoniEtAl05}; weaker polarization, $2-7\%$ on average, was found in MACS J0717.5 +3745; see \citet{BonafedeEtAl09}), and in one MH \citep[at a $10\%-20\%$ level, in A2390;][]{BacchiEtAl03}.
Spectral steepening with increasing $r$, increasing $\nu$, or
decreasing $T$, has been reported in several radio halos
\citep[][and references therein]{FerettiGiovannini08,FerrariEtAl08, GiovanniniEtAl09}; typically, $\alpha$ steepens from $-1.0$ to $-(1.3$--$1.5)$ in uncontaminated regions (see \S\ref{sec:Spectrum}).
A handful of GHs show a steep, $\alpha<-1.5$ spectrum, such as in A521 \citep[$\alpha=-1.86\pm0.08$;][]{DallacasaEtAl09} and in A697 \citep[$\alpha=-(1.7$--$1.8)$;][]{MacarioEtAl10}.

We analyze a sample of GHs and MHs using radio and X-ray data from the literature.
We find a universal correlation between the radio and X-ray surface brightness, which holds for both types of halos.
This correlation, combined for example with the observed dependence of the radio bright volume fraction upon cluster parameters \citep{CassanoEtAl07}, gives rise to the luminosity correlation known in GHs, and a similar correlation that we derive for MHs.
We determine the radio emissivity $j_\nu$ and its scaling with electron density and temperature $T$, using the surface brightness correlation and a model (XSPEC/MEKAL) for the X-ray emission.
Combined with other properties of GHs and MHs, this singles out secondary {\CRE} models with strong magnetic fields for both types of halos.
This generalizes the GH model of K09, and, considering the different environments of GHs and MHs, substantially strengthens it.
We propose that the cosmic-ray ions ({\CRIs}) are produced in supernovae, based on the halo spectra, the $j_\nu$ scaling, and a preliminary correlation between star formation and the radio to X-ray brightness ratio $\myeta$.
We show how the spectral and morphological properties of halos can be used to disentangle the distributions of {\CREs} and magnetic fields.
In particular, the magnetic field is gauged both by a radial break in $\myeta$ reflecting the onset of Compton losses, and by a spectral break in radio emission induced by diffractive p--p scattering.

The paper is arranged as follows.
In \S\ref{sec:Correlation} we show that GHs and MHs follow the same
correlation between specific radio power and \emph{coincident} X-ray
luminosity, suggesting that the different types of halos arise from similar processes.
A tight, universal correlation between the surface brightness in radio and in X-rays is presented in \S\ref{sec:RadialProfile}.
It is used to derive the scaling of
radio emissivity with the local electron number density $n$ and temperature
$T$, and to explain previous results for the average halo emissivity and the luminosity correlation.
We then show in \S\ref{sec:Theory} that
the inferred scaling favors a model in which the radio emission in
both GHs and MHs originates from secondary {\CREs} produced by
collisions of {\CRIs} with the intracluster gas,
synchrotron radiating most of their energy in strong magnetic fields.
We study the morphological and spectral
properties of the halos in \S\ref{sec:Model}, presenting new methods
for measuring the distributions of {\CRIs} and magnetic
fields. Finally, we discuss the origin of the {\CRIs} and
magnetic fields in \S\ref{sec:Discussion}, and summarize our results.
We assume a Hubble constant $H=70\km\se^{-1}\Mpc^{-1}$.
Error bars are $1\sigma$ confidence intervals.

\section{GHs and MHs: similar correlation between radio and coincident X-ray emission}
\label{sec:Correlation}

The 19 X-ray bright GH clusters in the \citet{BrunettiEtAl07} sample
follow the correlation
\begin{eqnarray} \label{eq:GH_Correlation}
(\nu P_{\nu,1.4})_{45} \simeq 10^{-4.2\pm 0.3} L_{X[0.1,2.4],45}^{1.7} \coma
\end{eqnarray}
where the subscript $45$ denotes units of $10^{45}\erg\se^{-1}$.
Here, $X[\epsilon_1,\epsilon_2]$ designates integration over photon
energies between $\epsilon_1$ and $\epsilon_2$ measured in keV (we use $X[0.1,2.4]$ unless otherwise stated), and
subscript $1.4$ means that the radio signal is evaluated at a
frequency $\nu=1.4\GHz$.  The normalization uncertainty in Eq.~(\ref{eq:GH_Correlation}) is dominated
by an intrinsic scatter among halos, needed to obtain an acceptable
fit (chosen as $\chi^2/N=1$ with $N$ being the number of degrees of
freedom).
Accounting for dispersion in $P_\nu$ leads to the correlation in the form of Eq.~(\ref{eq:GH_Correlation}), in agreement with K09, somewhat
flatter than in \citet{CassanoEtAl06}.

A sample of $6$ well-studied MHs had been reported
\citep{CassanoEtAl08} as being ``barely'' consistent with the GH
correlation of \citet{CassanoEtAl06}.  However, we
find that  this
sample agrees well with Eq.~(\ref{eq:GH_Correlation}) with $\chi^2/N=1.4$.
Nevertheless, within a partly overlapping sample of 6 MHs studied
recently \citep[][M09]{GovoniEtAl09}, the MHs
appear significantly underluminous with respect to the prediction of
Eq.~(\ref{eq:GH_Correlation}) after a careful removal (M09) of the
contamination from \AGNs.

These results do not imply that MHs are intrinsically fainter than
GHs. Rather, MHs have comparable radio power, if one corrects for
their smaller size. A natural generalization of
Eq.~(\ref{eq:GH_Correlation}) applicable for both GHs and MHs would
replace $L_X$ with the luminosity $\bar{L}_X$ of the radio bright region.
Following M09, we define the radio bright region based
on the surface brightness condition $I_\nu(r)/I_\nu(r=0)>e^{-3}\simeq 5\%$, with $r$
being the radial distance from the cluster center; equivalently, the
radius of this region is $R_\nu=3r_e$, with $r_e$ being the e-fold radius.

A quick way to estimate $\bar{L}_X$ is to correct $L_X$ using a model
for the X-ray brightness profile of the cluster.
Using an isothermal $\beta$-model \citep{CavaliereFusco76},
in which $n(r)=n_0 (1+r^2/r_c^2)^{-3\beta/2}$ for model parameters $\beta$, $r_c$ and $n_0$, we find that
\begin{eqnarray} \label{eq:GeneralizedLx}
\bar{L}_X = L_X \frac{\left(1+R_\nu^2/r_c^2\right)^{\frac{3}{2}-3\beta}-1} {\left(1+R_X^2/r_c^2\right)^{\frac{3}{2}-3\beta}-1} \fin
\end{eqnarray}
We use individual cluster $\beta$-model parameters from the literature; see Table \ref{tab:halos} for details.  The
radius $R_X$ of the X-ray region considered must be finite in order to ensure
convergence. Its choice is somewhat arbitrary; we use $R_X=12r_c$
so $R_\nu<R_X$ for all halos in the M09 sample.

\begin{sidewaystable*}[ht]{l}
%\begin{flushleft}
%\begin{table*}[ht]{l}
{
%\hfill{}
\vspace{1cm}
\caption{\label{tab:halos} Parameters of Halo Clusters in the Sample}
%\noindent\makebox[\textwidth]{%
% arXiv hide:
%{\fontsize{7}{7}\selectfont
\begin{tabular}{|ccccccccccccc|}%{|lllllllll|lll|}
\hline
%\hfill{}
%\tiny{ }&\tiny{ }&\tiny{ }&\tiny{ }&\tiny{ }&\tiny{ }&\tiny{ }&\tiny{ }&\tiny{ }&\tiny{ }&\tiny{ }&\tiny{ }\vspace{-2.5mm}\\
(1) & (2) & (3) & (4) & (5) & (6) & (7) & (8) & (9) & (10) & (11) & (12) & (13) \\
Cluster & Type & $z$ & $L_{X[0.1-2.4]}$ & $T$ & $r_e$ & $\langle j \rangle$ & $\beta$ & $r_c$ & $\bar{L}_{X[0.1-2.4]}$ & $B_{0,min}$ & $\langle\alpha\rangle$ & $Z(0.1r_{500})$ \\
%\cline{3-8}
\hline
% ---------------------- data from Mathematica --------------------
A401& GH & $0.074$ &  $6.3_{-0.1}^{+0.1}$  {(R02)} & $7.8_{-0.2}^{+0.2}$  {(A09a)} & $109_{-15}^{+21}$ & $4.1_{-1.0}^{+1.3}$ & $0.61_{-0.01}^{+0.01}$  {(C07)} & $175_{-7.1}^{+7.9}$  & $3.1_{-0.1}^{+0.1}$ & $7.5$ & $\alpha_{0.6}^{1.4}=-1.4$ (R81) & --- \\A545& GH & $0.154$ &  $5.7_{-0.5}^{+0.5}$  {(C06)} & $5.5_{-2.1}^{+6.2}$  {(D93)} & $150_{-11}^{+11}$ & $12.0_{-1.9}^{+2.2}$ & --- & --- & --- & --- & $\alpha_{1.4}<-1.4$ (G03) & --- \\A665& GH & $0.182$ &  $9.8_{-1.0}^{+1.0}$  {(C06)} & $7.9_{-0.2}^{+0.2}$  {(A09a)} & $236_{-15}^{+18}$ & $7.0_{-1.0}^{+1.0}$ & $0.54_{-0.01}^{+0.01}$  {(B06)} & $139_{-4.2}^{+4.2}$  & $7.1_{-1.0}^{+1.0}$ & $17.1$ & $\alpha_{0.3}^{1.4}=-1.04_{-0.02}^{0.02}$ (F04b) & $0.30\pm0.06$  \\A773& GH & $0.217$ &  $6.1_{-0.5}^{+0.5}$  {(B00)} & $7.4_{-0.3}^{+0.3}$  {(A09a)} & $111_{-10}^{+10}$ & $11.0_{-2.0}^{+2.4}$ & $0.61_{-0.01}^{+0.01}$  {(B06)} & $131_{-4.2}^{+4.2}$  & $3.7_{-0.5}^{+0.5}$ & $12.1$ & $\alpha_{0.3}^{1.4}=-1.02_{-0.26}^{+0.26}$ (K01) & $0.41\pm0.16$  \\A2163& GH & $0.203$ &  $17.1_{-0.3}^{+0.3}$  {(R02)} & $14.7_{-0.3}^{+0.3}$  {(V09)} & $394_{-7}^{+7}$ & $9.2_{-0.4}^{+0.4}$ & $0.80_{-0.03}^{+0.03}$  {(C07)} & $371_{-20.7}^{+21.4}$  & $15.2_{-0.3}^{+0.3}$ & $19.9$ & $\alpha_{0.07}^{1.4}=-1.02$ (V09) & $0.25\pm0.07$  \\A2218& GH & $0.176$ &  $4.6_{-0.2}^{+0.2}$  {(E98)} & $7.2_{-0.2}^{+0.2}$  {(A09a)} & $76_{-18}^{+26}$ & $20.0_{-9.0}^{+14.0}$ & $0.77_{-0.01}^{+0.01}$  {(B06)} & $190_{-4.3}^{+4.6}$  & $2.4_{-0.2}^{+0.2}$ & $7.5$ & $\alpha_{0.3}^{1.4}=-1.46_{-1.0}^{1.0}$ (K01) & $0.37\pm0.15$  \\A2219& GH & $0.226$ &  $12.7_{-1.0}^{+1.0}$  {(C06)} & $9.5_{-0.4}^{+0.6}$  {(C06)} & $359_{-14}^{+18}$ & $5.4_{-0.6}^{+0.6}$ & $0.66_{-0.01}^{+0.07}$  {(F04)} & $208_{-23.8}^{+99.1}$  & $11.2_{-1.0}^{+1.0}$ & $25.3$ & $\alpha_{0.3}^{1.4}=-0.9_{-0.1}^{+0.1}$ (O07) & --- \\A2254& GH & $0.178$ &  $4.3_{-0.4}^{+0.4}$  {(B00)} & $7.5_{-1.5}^{+0.5}$  {(C06)} & $238_{-39}^{+57}$ & $9.7_{-2.8}^{+3.4}$ & --- & --- & --- & --- & $\alpha_{1.4}^{1.7}\sim -1.2$ (G01) & --- \\A2255& GH & $0.081$ &  $2.6_{-0.1}^{+0.1}$  {(C06)} & $6.9_{-0.1}^{+0.1}$  {(A09a)} & $203_{-6}^{+6}$ & $3.4_{-0.2}^{+0.2}$ & $0.80_{-0.03}^{+0.03}$  {(C07)} & $424_{-22.9}^{+25.0}$  & $1.7_{-0.1}^{+0.1}$ & $7.4$ & $\alpha_{0.3}^{1.4}=-2.06_{-0.57}^{+0.57}$ (K01) & --- \\A2319& GH & $0.056$ &  $8.2_{-0.1}^{+0.1}$  {(C07)} & $8.8_{-0.2}^{+0.3}$  {(C06)} & $198_{-6}^{+7}$ & $5.4_{-0.4}^{+0.4}$ & $0.59_{-0.01}^{+0.01}$  {(C07)} & $203_{-9.3}^{+10.0}$  & $5.1_{-0.1}^{+0.1}$ & $3.8$ & $\alpha_{0.3}^{1.4}=-1.28_{-0.35}^{+0.35}$ (K01) & $0.31\pm0.04$  \\A2744& GH & $0.308$ &  $13.1_{-2.4}^{+2.4}$  {(C06)} & $8.6_{-0.3}^{+0.4}$  {(C06)} & $275_{-9}^{+9}$ & $25.0_{-1.6}^{+1.7}$ & $0.57_{-0.06}^{+0.12}$  {(F04)} & $251_{-2.4}^{+120}$  & $8.1_{-2.4}^{+2.4}$ & $16.0$ & $\alpha_{0.07}^{1.4}=-1.1$ (V09) & --- \\RXJ1314& GH & $0.244$ &  $7.3_{-0.7}^{+0.7}$  {(M01)} & $8.7_{-0.4}^{+0.4}$  {(M01)} & $160_{-38}^{+61}$ & $12.0_{-5.0}^{+8.0}$ & $0.77_{-0.23}^{+0.23}$  {(V02)} & $286_{-100}^{+100}$  & $4.9_{-0.7}^{+0.7}$ & $10.9$ & $\alpha_{0.07}^{1.4}=-1.4$ (V09) & --- \\\hline
A1835& MH & $0.253$ &  $16.3_{-1.2}^{+1.2}$  {(B00)} & $7.1_{-0.1}^{+0.1}$  {(A09a)} & $102_{-31}^{+70}$ & $33.0_{-13.0}^{+22.0}$ & $0.53_{0.00}^{+0.00}$  {(S08)} & $29.0_{-0.3}^{+0.3}$  & $15.6_{-1.2}^{+1.2}$ & $33.1$ & --- & $0.28\pm0.05$  \\A2029& MH & $0.076$ &  $8.5_{-0.1}^{+0.1}$  {(C07)} & $6.9_{-0.1}^{+0.1}$  {(A09a)} & $53_{-6}^{+6}$ & $54.0_{-13.0}^{+19.0}$ & $0.58_{0.00}^{+0.00}$  {(C07)} & $59.3_{-1.4}^{+1.4}$  & $4.9_{-0.1}^{+0.1}$ & $8.3$ & $\alpha_{0.2}\simeq -1.35$ (S83) & $0.38\pm0.06$  \\A2390& MH & $0.228$ &  $15.8_{-0.2}^{+0.2}$  {(A03)} & $9.3_{-0.1}^{+0.1}$  {(B07)} & $36_{-4}^{+4}$ & $3100_{-800}^{+1000}$ & $0.47_{0.00}^{+0.00}$  {(S08)} & $46.0_{-1.0}^{+1.0}$  & $5.3_{-0.2}^{+0.2}$ & $9.5$ & $\alpha_{0.07}^{0.4}\simeq -1.10$ (A06) & --- \\Ophiuchus& MH & $0.028$ &  $6.1_{-0.2}^{+0.2}$  {(C07)} & $10.2_{-0.4}^{+0.3}$  {(C07)} & $105_{-11}^{+13}$ & $4.7_{-0.8}^{+0.9}$ & $0.75_{-0.03}^{+0.04}$  {(C07)} & $199_{-15.0}^{+16.4}$  & $3.8_{-0.2}^{+0.2}$ & $6.9$ & --- & --- \\Perseus& MH & $0.018$ &  $8.2_{-0.1}^{+0.1}$  {(C07)} & $6.4_{-0.1}^{+0.1}$  {(F04)} & $23_{-1}^{+1}$ & $3600_{-500}^{+500}$ & $0.54_{0.00}^{+0.01}$  {(C07)} & $45.0_{-0.7}^{+1.4}$  & $2.5_{-0.1}^{+0.1}$ & $4.8$ & $\alpha_{0.3}^{1.5}\sim -1.2$ (G04) & $0.42\pm0.01$  \\RXJ1347& MH & $0.451$ &  $27.5_{-1.1}^{+1.1}$  {(F04)} & $10.7_{-0.1}^{+0.1}$  {(A09b)} & $52_{-11}^{+17}$ & $1800_{-500}^{+700}$ & $0.54_{0.00}^{+0.00}$  {(B06)} & $23.3_{-0.5}^{+0.5}$  & $22.6_{-1.1}^{+1.1}$ & $32.4$ & --- & $0.39\pm0.04$  \\
% ---------------------- data from Mathematica --------------------
\hline
\end{tabular}
% arXiv hide
%}%close font
\hfill{}
\vspace{0.2cm}
\\
{\bf Columns}: (1) cluster name (1RXS J131423 .6-251521 abbreviated RXJ1314; RXJ1347.5-1145 abbrev. RXJ1347); (2) halo type (GH or MH); (3) redshift $z$; (4) X-ray luminosity between $0.1$ and $2.4\keV$, in units of $10^{44}\erg\se^{-1}$; (5) temperature in $\keV$; (6) e-fold radius of radio brightness in kpc (M09); (7) average $1.4\GHz$ specific emissivity in units of $10^{-43}\erg\se^{-1}\cm^{-3}\Hz^{-1}$ (M09); (8-9) exponent $\beta$ and core radius $r_c$ in kpc for an isothermal  $\beta$-model of the cluster; (10) X-ray luminosity of the radio bright region (see \S\ref{sec:Correlation}) between $0.1$ and $2.4\keV$, in units of $10^{44}\erg\se^{-1}$; (11) minimal central magnetic field $B_{0,min}$ in units of $\mu$G, assuming the $\beta$-model and that $B(r=\min\{r_b,3r_e\})=B_{cmb}$ (corresponding to $f_{ic}=0.5$ in Eq.~\ref{eq:B0}), where the radial break radius $r_b=55\kpc$ in Perseus, $135\kpc$ in A2029, $60\kpc$ in A2319, and $r_b>3r_e$ in the other halos; (12) average spectral index $\alpha_{\nu_1}^{\nu_2}$ between frequencies (in GHz) $\nu_1$ and $\nu_2$, or $\alpha_{\nu}$ around frequency $\nu$; (13) Metallicity measured at $r=0.1r_{500}$ in solar units \citep{SnowdenEtAl08}.
\\ \\
{\bf References}: References are given in parentheses. For conflicting references, we adopt the tighter estimate if applicable, and otherwise use the most recent result.
A03: \citet{AllenEtAl03};
A06: \citet{AugustoEtAl06};
A09a: \citet{AnderssonEtAl09};
A09b: \citet{AndersonEtAl09};
B00: \citet{BohrinferEtAl00};
B03: \citet{BacchiEtAl03};
B06: \citet{Bonamente06};
B07: \citet{BaldiEtAl07};
C06: \citet{CassanoEtAl06};
C07: \citet{ChenEtAl07};
D93: \citet{DavidEtAl93};
F04: \citet{FukazawaEtAl04};
F04b: \citet{FerettiEtAl04b}
F07: \citet{FerettiEtAl97};
G01: \citet{GovoniEtAl01B};
G03: \citet{GiovanniniEtAl03};
G04: \citet{GittiEtAl04};
H80: \citet{HarrisEtAl80};
K01: \citet{KempnerSarazin01};
M00: \citet{MatsumotoEtAl00};
M01: \citet{MatsumotoEtAl01};
M07: \citet{MorandiEtAl07};
R81: \citet{RolandEtAl81}
R02: \citet{ReiprichBohringer02};
S83: \citet{SleeSiegman83};
S08: \citet{SantosEtAl08};
S09: \citet{SandersonEtAl09};
V02: \citet{ValtchanovEtAl02};
V09: \citet{vanWeerenEtAl09};
WB03: \citet{WorrallBirkinshaw03};
W00: \citet{White00}. \\
}%close makebox
\vspace{0.5cm}
%\end{table*}
%\end{flushleft}
\end{sidewaystable*}

\begin{figure}[h]
\centerline{\epsfxsize=8.5cm \epsfbox{\myfig{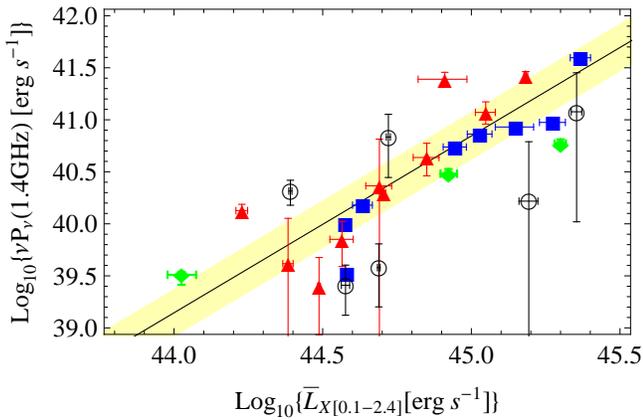}}}
\caption{ Radio power $\nu P_\nu$ at $1.4\GHz$ of GHs (red triangles) and MHs (black circles) from
M09, plotted against the 0.1-2.4 keV X-ray luminosity $\bar{L}_{X}$ of the radio
bright region (see \S\ref{sec:Correlation} and table \ref{tab:halos} for details).
The total X-ray luminosity $L_{X}$ is used in cases where $\bar{L}_X$ could
not be determined, for GHs \citep[blue squares, from][]{CassanoEtAl06} and MHs \citep[green diamonds, from][]{CassanoEtAl08}.  The solid
line and shaded region show the best fit and $1\sigma$ interval of the GH correlation in
Eq.~(\ref{eq:GH_Correlation}).
\label{fig:MH_CorrectedCorrelation}
}
\end{figure}

Figure \ref{fig:MH_CorrectedCorrelation} shows various GHs and MHs in
the resulting $\bar{L}_X-\nu P_\nu$ plane.  Replacing $L_X$ by
$\bar{L}_X$ results in slightly better agreement of the M09 GHs with
Eq.~(\ref{eq:GH_Correlation}), yielding $\chi^2/N=1.3$ instead of 1.6
(before propagating $\beta$ model uncertainties).  However, the MH
agreement becomes much better, $\chi^2/N=2.4$ instead of 3.7.  Note
that the fit is not expected to be as good for MHs as it is for GHs,
because a $\beta$-model is less appropriate for a CC.  These results
suggest that the relation between radio and X-ray emission is similar in GHs and in MHs.
For more accurate, model-independent results, we next examine a more local measure of the emission, and consider the radio and X-ray morphologies.

\section{Correlation between Radio and X-ray surface brightness: universal emissivity}
\label{sec:RadialProfile}

A more useful manifestation of the radio-X-ray correlation is the morphological similarity between radio and X-ray emission.
A linear correlation between the radio surface brightness $I_\nu$ and the X-ray brightness
$F_X$ was found by \citet{GovoniEtAl01} in two GH clusters (A2744 and A2255, inspected individually),
while a sublinear power law was found in two other GHs (A2319 and Coma).

\begin{figure}[h]
\centerline{\epsfxsize=9.5cm \epsfbox{\myfig{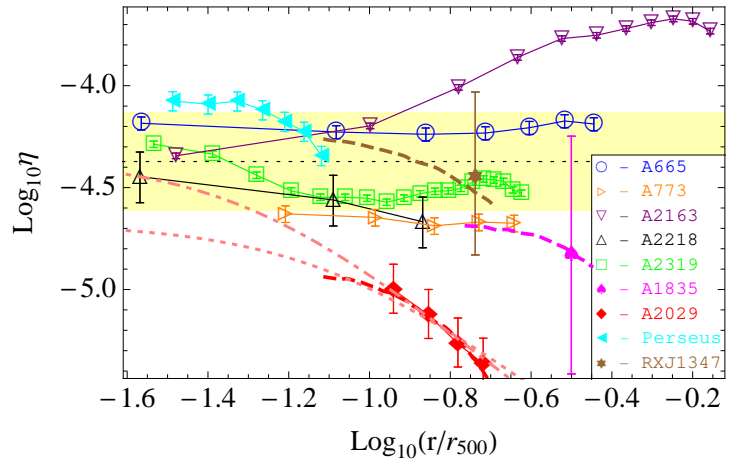}}}
\caption{ Radial profiles of the radio-to-X-ray ratio $\myeta$ (defined in Eq.~(\ref{eq:eta_def})) in the GHs
(open symbols) and MHs (filled symbols) with overlapping profiles from VLA (M09) and from \emph{XMM-Newton} \citep[][k-corrected]{SnowdenEtAl08}.
The MH data is shown only in regions where the AGN contamination is $<10\%$;
this is supplemented by the exponential model fits (dashed curves) given by M09, plotted in the range where radio detection is $>3\sigma$ and the MH contribution is $>30\%$.
Error bars are the $1\sigma$ confidence intervals of the radio
normalization $I_0$ (M09), and the solid lines serve to guide the eye.
The $1\sigma$ best fit of Eq.~(\ref{eq:local_correlation}) is shown as
a thin dashed line in a shaded region.
Model fits for A2029 (dotted and dot-dashed curves; see \S\ref{sec:Morphology}) are also shown.
\label{fig:MH_RadialTCorrectedProfiles}
}
\end{figure}

We investigate the connection between the radio and the X-ray surface brightness in GH
and MH clusters by examining the radial brightness profiles published in the literature.
We consider not only the radio to X-ray relation within each cluster, but also
compare the ratio between $\nu I_\nu$ and $F_x$ in different clusters.
Thus, Figure \ref{fig:MH_RadialTCorrectedProfiles} shows the radial profile of the dimensionless ratio
\begin{eqnarray} \label{eq:eta_def}
\myeta(r) \equiv \frac{\nu I_\nu(1.4\GHz)}{F_{X[0.1-2.4]}}
\end{eqnarray}
between radio and X-ray
surface brightness, for all six GHs and four MHs with published radial profiles from both Very Large Array (VLA; M09) and \emph{XMM-Newton} \citep{SnowdenEtAl08} observations.
Distances are normalized to $r_{500}$, the radius
enclosing $500$ times the critical density of the Universe
\citep[calculated using the best fit of][]{ZhangEtAl08}.

All halo clusters except the MH in A2029 converge on a similar value of $\myeta$ at small
radii.  The innermost GH data is best fit by
\begin{eqnarray} \label{eq:local_correlation}
\myeta_0 \equiv \myeta(r=0) = 10^{-4.4\pm0.2} \coma % 10^{-4.37\pm0.24} \coma
\end{eqnarray}
where the uncertainty is again larger than the measurement errors and mostly attributed to intrinsic scatter between halos.
The inner MH data also lie within the $1\sigma$ scatter, except A2029 which is significantly fainter in radio (but is observed only where $\myeta(r)$ rapidly declines).
Therefore, Eq.~(\ref{eq:local_correlation}) appears to hold at small
radii for both GHs and MHs.
A possible explanation for the relatively high radio brightness of Perseus and A665, and the low brightness of A2029, is discussed in \S\ref{sec:CRI_Origin}.
As in \S\ref{sec:Correlation}, we used only GHs, which are better understood than MHs, to derive the correlation Eq.~(\ref{eq:local_correlation}).
If, instead, we use the innermost data of both GHs and MHs, we obtain a broader best fit, $\eta_0 \simeq 10^{-4.4\pm0.4}$, % $\eta \simeq 10^{-4.41\pm0.37}$.
due to the bright MH in Perseus and the faint MH in A2029.

Although there is some scatter among different clusters, Fig.~\ref{fig:MH_RadialTCorrectedProfiles} shows that $\myeta$ is
fairly uniform within each GH.
An exception is A2163, in which $\myeta(r)$ monotonically increases out to $r\simeq 0.6r_{500} \simeq 1200\kpc$.
However, this is the only cluster shown in Fig.~\ref{fig:MH_RadialTCorrectedProfiles} that harbors a radio relic \citep{FerettiEtAl01}, which dominates the emission near the $\myeta$ maximum (M09, figure 3). This suggests that the peculiar $\myeta$ profile in A2163 is associated with the radio relic.
A peculiar $\myeta(r)$ pattern is seen in A2319, with a minimum at $r\simeq 0.1r_{500} \simeq 150\kpc$. Interestingly, this cluster shows both GH and MH characteristics, harboring both a CF at $r\simeq 100$--$200\kpc$ and a subcluster at $r\simeq 500\kpc$ \citep{GovoniEtAl04}.
In \S\ref{sec:Morphology} we argue that the clumpy radio morphology of this cluster suggests weak magnetization, enhanced at both small and large radii in an intermediate phase between a GH and a MH.
The declining $\myeta(r)$ profiles of the MHs are discussed in \S\ref{sec:Morphology}.

In order to derive the radio emissivity from the brightness correlation Eq.~(\ref{eq:local_correlation}), we must model the X-ray emission.
The X-ray emissivity calculated using the MEKAL model
\citep{MeweEtAl85, MeweEtAl86, Kaastra92, LiedahlEtAl95} in XSPEC v.12.5 \citep{Arnaud96} is
nearly independent of temperature and metallicity.
It is given by
\begin{eqnarray} \label{eq:FxMEKAL}
j_{X[0.1-2.4]}\simeq 8.6\times 10^{-28}\frac{n_{-2}^2 Z_{0.3}^{0.04}}{T_{10}^{0.1}}\erg\se^{-1}\cm^{-3} \coma
\end{eqnarray}
where $n_{-2}$ is the electron number density $n$ in units of
$10^{-2}\cm^{-3}$,
$T_{10}\equiv(k_BT/10\keV)$ with $k_B$ being Boltzmann's constant,
and $Z_{0.3}$ is the metallicity $Z$ in units of $0.3 Z_\odot$.

Fairly uniform values of $\myeta$ are found within each cluster (excluding the outer regions of A2163), and similar values are found among different halos, in regions spanning two orders of magnitude in density.
This implies that the radio emissivity $j_\nu$, like $j_X$, scales as the gas density squared.
Indeed, M09 recently found (in their section 5.1) that assuming $j_\nu\propto n^2$ reproduces the radial profiles of some GHs and MHs.
We conclude that the linear surface brightness relation Eq.~(\ref{eq:local_correlation}) reflects a similar, local relation between the radio and X-ray emissivities,
\begin{eqnarray} \label{eq:emissivity_relation}
\nu j_\nu \simeq \eta_0  j_X \fin
\end{eqnarray}

We examine the temperature dependence of the radio emissivity within a given cluster by fitting $\myeta$ in uncontaminated GHs as a power law, such that $j_\nu\propto n^2 (T/T_0)^\mypT$, with $T_0$ being the central temperature and $\mypT$ being a free parameter.
This yields $\mypT=0.2\pm 0.5$, consistent with weak or no temperature dependence within the cluster.
Combining this with the above results Eqs.~(\ref{eq:local_correlation})-(\ref{eq:emissivity_relation}), we conclude that the synchrotron emissivity in halos is well fit by
\begin{eqnarray} \label{eq:nu_j_nu}
\nu j_\nu \simeq 10^{-31.4\pm0.2} n_{-2}^2 (T/T_0)^{0.2\pm0.5} \erg\se^{-1}\cm^{-3} \coma
\end{eqnarray}
at least near the centers of the halos.

The scatter in Eqs.~(\ref{eq:local_correlation}) and (\ref{eq:nu_j_nu}) indicates that the radio emissivity is not a function of the local density and temperature alone.
The dispersion in $j_\nu$ probably reflects dependence upon additional cluster properties, such as the cluster's mass and star formation history. It has standard deviation $\sigma(\log j_\nu)\simeq\log(1.6)$ among the GHs in our sample, or $\sigma(\log j_\nu)\simeq\log(2.5)$ if we include the MHs.
This relatively small dispersion explains the similar
volume-averaged emissivity $\ave{j_\nu}$ of different GHs, and the
high level of variation in $\ave{j_\nu}$ among MHs, where $\ave{j_\nu}$ is sometimes
two orders of magnitude larger than in GHs, as reported by
\citet{CassanoEtAl08} and M09; see Table \ref{tab:halos}.  Indeed, the GHs in their sample have
comparable sizes and central densities $n_0\simeq 10^{-2}\cm^{-3}$,
whereas the MHs have variable scale and $n_0>10^{-1}\cm^{-3}$ in some
cases.

We have derived linear relations between the radio and X-ray emissivity (Eq.~\ref{eq:emissivity_relation}) and surface brightness (Eqs.~\ref{eq:eta_def} and \ref{eq:local_correlation}) of halos, but a superlinear, $P_\nu\propto L_x^{1.7}$ relation between the integrated luminosity in the two bands (Eq.~\ref{eq:GH_Correlation}).
This indicates that, in addition to the $j_\nu\propto n^2(T/T_0)^\kappa$ dependence, the total radio power of a halo must further increase, on average, with the mass and temperature of the cluster.
Such a behavior could result in part from a (direct or indirect) mass or temperature dependence of $j_\nu$.
However, the dispersion in $j_\nu$ is small, and we find no evidence for such a correlation.
The effect is probably dominated by the different scalings of the radio and X-ray bright volumes with the cluster parameters, as we qualitatively show next.

The length scale $R_\nu$ of GHs was found to depend superlinearly on the virial radius $R_{vir}$, $R_\nu\sim R_{vir}^{\lambda_R}$, with $\lambda_R = 2.6\pm0.5$ \citep{CassanoEtAl07}.
In order to estimate the effect of this scaling on the $P_{\nu}$--$L_X$ correlation, we crudely approximate the luminosity within a sphere of radius $R$ as a power-law, $L(<R)\propto R^{\lambda_L}$.
In a $\beta$-model, this is a poor approximation, with power-law indices $0.5<\lambda_L<2$ in the parameter range $0.5<\beta<0.7$, $1<r/r_c<4$, relevant to GHs.
For simplicity, we assume that $L_X$ reflects the emission within a sphere of radius $\propto  R_{vir}$.
For constant $\myeta$, we may then write
\begin{eqnarray} \label{eq:RoughScalings}
\frac{P_{\nu}}{L_{X}} \propto \left( \frac{R_\nu}{R_{vir}} \right)^{\lambda_L} \propto R_{vir}^{(\lambda_R - 1)\lambda_L} \propto L_X^{0.3(\lambda_R-1)\lambda_L} \coma
\end{eqnarray}
where we used the observed scalings $R_{vir}\sim T^{0.6}$ \citep{ZhangEtAl08} and $L_{X}\sim T^2$ \citep{Markevitch98} in the last proportionality (recall that $X$ represents the energy range $0.1$--$2.4\keV$).
Equation (\ref{eq:RoughScalings}) yields $P_{\nu}\propto L_X^{(1.2-2.0)}$, where we accounted for the $\lambda_L$ range but have not propagated the scaling errors.
While this demonstrates that the $R_\nu$--$R_{vir}$ scaling may suffice to reconcile the nonlinear luminosity relation Eq.~(\ref{eq:GH_Correlation}) with the linear emissivity relation Eq.~(\ref{eq:emissivity_relation}), much better modeling is required in order to identify all factors governing the luminosity correlation.
For example, we neglected halo asymmetry and possible variations in $\myeta$ at large radii (as in A2163).

\section{Universal Radio Mechanism}
\label{sec:Theory}

In the preceding sections we examined the luminosity and surface brightness properties of GHs and MHs, in radio and in X-rays, and derived the radio emissivity, without making any assumptions regarding the particles and magnetic fields responsible for the radio emission.
Here we explore the implications of the radio-X-ray correlation and the radio scalings derived above.
In \S\ref{sec:ScalingImplications}, we discuss various models for radio halos, and show that only models with strong magnetic fields and secondary {\CREs} are consistent with the observations.
In \S\ref{sec:TemporalEvolution} we briefly discuss the effect of time dependent magnetic fields.

\subsection{Secondary {\CREs}, strong magnetic fields}
\label{sec:ScalingImplications}

The spectral slope near the center of most radio halos is
$\alpha\simeq -1$ (see \S \ref{sec:Spectrum}).
A synchrotron spectrum of this type is emitted by rapidly cooling
{\CREs}, injected with approximately constant energy per decade in particle energy $E_e$.
The synchrotron emissivity corresponding to a logarithmic {\CRE} energy
density injection rate $Q\equiv du_e/(dt\,d\ln E_e)=\constant$ is
given by
\begin{eqnarray} \label{eq:injection_definition}
\nu j_\nu \equiv \nu \frac{du_{\gamma}}{dt\,d\nu} = \frac{Q/2}{1+(B_{cmb}/B)^{2}} \coma
\end{eqnarray}
where the denominator accounts for Compton losses, $B_{cmb}\equiv
(8\pi u_{cmb})^{1/2}\simeq 3.2(1+z)^2\muG$ is the amplitude of the magnetic field which has the same
energy density as the CMB, and $u_j$ is the energy density of component $j$ ($\gamma$ for radio photons, $e$ for {\CREs}, etc.).
In the $B\gg B_{cmb}$ limit, the synchrotron emissivity Eq.~(\ref{eq:nu_j_nu}) extracted from the observations yields a direct estimate of the {\CRE} injection rate,
\begin{eqnarray} \label{eq:injection}
Q \simeq \frac{Q}{1+(B_{cmb}/B)^{2}} = 10^{-31.1\pm0.2} n_{-2}^2 \erg\se^{-1}\cm^{-3} \fin \,\,\,\,\,
\end{eqnarray}
As {\CREs} do not have time to diffuse before they lose their energy,
this result applies locally. It holds in both GHs and MHs.

The remarkably small scatter found in the radio-X-ray correlations among
different GHs has been interpreted as implying a robust radio
mechanism that keeps the synchrotron emissivity
narrowly distributed around a universal function of the plasma density
and perhaps also temperature.  Equation~(\ref{eq:injection}) and the
narrow scatter in its normalization suggest that the universal
quantity in radio halos --- both GHs and MHs --- is $Q/n^2$.

Our results strongly emphasize the robustness of the radio emission mechanism.
First, we find a tight radio-X-ray correlation not only in GHs but
also in MHs.  As the latter have physical properties that are considerably
different from GHs (smaller size, higher ambient density, nearby AGN,
CF association), the radio emission mechanism must be sufficiently
robust to reproduce the same levels of $Q/n^2$ over a wide range of
physical conditions.  Second, we confirm that a tight radio-X-ray
correlation exists not only in the total cluster luminosity but also
in the local surface brightness, and that the ratio $\myeta$ between radio and X-ray brightness is fairly uniform within each cluster.  This is particularly striking in MHs, where the coincident density and temperature profiles are steep.

K09 have pointed out that in the strong magnetization regime, defined as
$B>B_{cmb}$, radio emission from GHs is independent of the precise value of
$B$, and so the tight correlation only constrains the {\CRE} injection $Q$.
In contrast, in the weak magnetization regime $B<B_{cmb}$, the product
$QB^2$ should be universal, requiring a physical mechanism that
carefully balances {\CRE} injection and magnetic fields both with each
other and with the ambient gas.  Furthermore, in a strongly magnetized halo model,
the transition from a constant to a rapidly declining ($\propto B^2$) behavior as $B$ drops below
$B_{cmb}$ naturally explains the GH bimodality observed (K09).

The similar values of $\myeta$ we find in GHs and in MHs strengthen this argument considerably, because it is difficult to come up with a double feedback mechanism ({\CRE}--magnetic fields--ambient plasma) that operates identically in the different environments of GHs and MHs, without fine tuning.
The bimodality argument does not apply to MHs, however, as their distribution has not been shown to be bimodal, and may in fact be continuous if high magnetization is ubiquitous in CC centers (see discussion in \S\ref{sec:Magnetization}).

Two types of models have been proposed for {\CRE} injection: {\it (i)}
secondary production by hadronic collisions involving {\CRIs}
\citep{Dennison80}; and {\it (ii)} in-situ turbulent acceleration or
reacceleration of primary {\CREs} \citep{EnsslinEtAl99}.  These models typically assume fixed ratios between the energy densities of the primary
particles (either {\CRIs} or {\CREs}), magnetic fields, and thermal
plasma $u_{th}\sim nT$.  Other possibilities involve a primary {\CRI} distribution that has energy density $u_i\propto n$ (such a scaling is less likely for the magnetic
fields or the rapidly cooling {\CREs}), or a magnetic field frozen
into the plasma, $B\propto n^{2/3}$.
The synchrotron emissivity in each of
the nine model variants corresponding to these three primary
distributions folded with different magnetization levels and scalings, is
shown in Table \ref{tab:models}.

\begin{table}[h]
\begin{center}
\caption{\label{tab:models}
% ApJ hide "\\":
Radio emissivity in model variants with different\\ distributions of {\CREs} and magnetic fields}
%\hspace{-10mm}
\begin{tabular}{|l|ccc|}
\hline
\,Primary scaling $\rightarrow$ & {\CREs} & {\CRIs} & {\CRIs} \\
$\downarrow$ Magnetic scaling\, & $u_e\propto nT$  & $u_i\propto nT$ & $u_i\propto n$ \\
\hline
$B_{cmb}>B\propto \sqrt{nT}$  & $j_\nu\propto n^2T^2$ & $j_\nu\propto n^3T^2$ & $j_\nu\propto n^3T$ \\
$B_{cmb}>B\propto n^{2/3}$  & $j_\nu\propto n^{7/3}T$ & $j_\nu\propto n^{10/3}T$ & $j_\nu\propto n^{10/3}$ \\
$B>B_{cmb}$  & $j_\nu\propto nT$ & $\bm{j_\nu\propto n^2T}$ & $\bm{j_\nu\propto n^2}$ \\
\hline
\end{tabular}
\end{center}
\end{table}

Among these model variants, only the two secondary {\CRE} models in which the magnetic field is
strong (models highlighted as boldface in the table) are consistent with
the scaling Eq.~(\ref{eq:nu_j_nu}) --- $j_\nu\propto n^2 T^{\mypT}$ where $\mypT=0.2\pm0.5$ --- and with Eq.~(\ref{eq:injection}).
Both models are consistent with our data.
A slightly better fit to the GH profiles is obtained with $u_i\propto n$ (\ie $\mypT= 0$; see \S\ref{sec:RadialProfile}), but more data is needed in order to establish the thermal dependence of the {\CRI} distribution with sufficient statistical significance.

An independent argument in favor of these two models is the environment of MHs.
MHs are found in relaxed clusters such as A2029, where only CFs reveal
deviations from hydrostatic equilibrium.  Such CFs reflect subsonic
bulk shear flows and strong magnetization, but were not associated
with particle acceleration.  The CF-MH connection therefore supports
both secondary {CREs} and strong magnetization in MHs.
The similarity in the values of $\myeta$ we find in GHs and in MHs implies, transitively, that the same holds for GHs.

Additional evidence supporting the presence of strong magnetic fields and the absence of primary {\CREs} is found by examining the morphological and spectral properties of halos, as discussed in \S\ref{sec:Model}.
Notice, for example, that the brightness correlation shown in Fig.~(\ref{fig:MH_RadialTCorrectedProfiles}) is strongest in the centers of halos, but diminishes at larger radii, where turbulent activity associated with mergers is expected and where merger shocks and relics are found. In primary {\CRE} models, one would not expect the correlation to preferentially tighten \emph{away} from the turbulent regions.

\subsection{Temporal variations}
\label{sec:TemporalEvolution}

The preceding discussion is strictly valid only as long as the timescale for substantial changes in the magnetic field is longer than the cooling time of the {\CREs}, $t_{cool}$.
For {\CREs} that emit synchrotron radiation received with characteristic frequency $\nu$,
\begin{eqnarray}
t_{cool} \simeq 0.13 \left[\frac{4\left(\frac{B\sqrt{3}}{B_{cmb}}\right)^{-3/2}}{1+\left(\frac{B}{B_{cmb}}\right)^{-2}}\right] \nu_{1.4}^{-\frac{1}{2}}(1+z)^{-\frac{7}{2}} \Gyr \coma
\end{eqnarray}
where the term in square brackets peaks at unity when $B=B_{cmb}/\sqrt{3}$, and scales as $B^{-3/2}$ for $B\gg B_{cmb}$.
In GHs, $t_{cool}$ is much shorter than the $\gtrsim 1\Gyr$ timescale characteristic of the halo lifetime (K09).

However, significant local variations in the magnetic field can take place on a shorter timescale, of the order of the sound crossing time of the turbulent eddies, $t_s\sim l/c_s$.
This is shorter than $t_{cool}$ for a sound velocity $c_s\sim 10^3\km\se^{-1}$ and eddy length scales $l\lesssim 100\kpc$.
(Note that substantial magnetic power is measured on coherence scales $\sim 10\kpc$; see \S\ref{sec:Magnetization}.)
The radio emission should therefore be averaged over $t_{cool}$ and over the beam.
Nevertheless, as long as many eddies contribute to the emission, and the variations in magnetic energy density remain of order unity, this correction would be small.
Notice that fast changes in magnetic configuration over a \emph{light} crossing time, if present, may be observed as temporal radio variations by next generation telescopes such as the Square Kilometre Array (SKA).
Indeed, the milli-arcsecond resolution attainable by SKA \citep{SchilizziEtAl07} corresponds to a light-crossing time of less than a year, for nearby halos at redshift $z\lesssim 0.02$.

In MHs, the magnetic fields are probably associated with sloshing activity in the core (see \S\ref{sec:Magnetization}).
The characteristic timescale for the decay of core sloshing is $\gtrsim 1\Gyr$ \citep[\eg][]{AscasibarMarkevitch06}, much longer than $t_{cool}$.
The timescale for the buildup of sloshing depends on its trigger mechanism; in a merger induced scenario this is again $\gtrsim 1\Gyr$.
However, local variations in the magnetic field could occur over the radial sound crossing time, $t_s\sim r/c_s$, or on the crossing time of the magnetic structures associated for example with CFs.
These timescales could be shorter than $t_{cool}$ in the central $\sim 50\kpc$ (note that in these regions $B$ usually significantly exceeds $B_{cmb}$; see \S\ref{sec:Morphology}).
As in GHs, averaging over $t_{cool}$ and over the beam could introduce small variations in radio brightness, in particular at the edges of MHs.

In both GHs and MHs, $Q/n^2$ is weakly sensitive to variations in {\CRE} injection, through changes in the fractional energy of the {\CRIs}, $u_i/u_{th}$.
Thus, $Q/n^2$ should be replaced by its value averaged over $t_{cool}$ and over the beam.
This correction should be very small, except near a cosmic-ray source.

\section{Signature of secondary {\CREs} in strong fields}
\label{sec:Model}

In the preceding sections we showed that observations support a halo model which invokes secondary {\CREs} in strong magnetic fields as the origin of radio halos, both GHs and MHs.
In such a model, the radio emissivity $j_\nu \propto n_{p}n B^2/(B^2+B_{cmb}^2)$ depends weakly on $B$.
This model recovers the observed radio-X-ray correlations, provided that the local energy fraction of {\CRPs} is narrowly distributed about a universal, weak function of the cluster parameters, because then $j_\nu\sim n^2\sim j_X$.

Such a model entails particular morphological and spectral properties of GHs and MHs.
Utilizing the model, we show how these properties can be used to shed light on
halo observations, test the model, and gauge its parameters.
In particular, we demonstrate how the distributions of {CREs} and magnetic fields can be
measured separately, rather than their degenerate product as done in most other models.

\subsection{Morphology: $B<B_{cmb}$ radio suppression}
\label{sec:Morphology}

In our model, both GHs and MHs are regions in which strong magnetic
fields with $B\gtrsim B_{cmb}$ indirectly illuminate the cluster's {\CRI} population in
radio waves.
This explains the spatial coincidence between MH edges and CFs, which are present in more than half of all CCs.
Bulk shear flow is believed to magnetize the plasma across and beneath the CFs; there is however no evidence for shear above the CFs \citep{KeshetEtAl10}.
Therefore, observations of sharp MH termination coincident with CFs reflect the transition from strong to weak magnetic fields.
This is probably the reason for the rapid, nearly
exponential $\myeta(r)$ cutoff in Perseus above $r\simeq 0.05r_{500}\simeq 50\kpc$, shown in
Fig. \ref{fig:MH_RadialTCorrectedProfiles}.  Note that CFs are not
spherical; the radial decay may result from one or several CFs extending
over various radii, seen projected and radially binned.

In GHs, and in MHs away from CFs, it is natural to assume that the magnetic field decays
with $r$, possibly as some fixed fraction of equipartition,
$B^2\propto nT$.  At some distance $r_b$, $B$ drops below $B_{cmb}$,
leading to a $\sim(B/B_{cmb})^2$ suppression in the radio emissivity.
This can produce a radial break in the projected radio profile, with
$\myeta\simeq \constant$ at $r\ll r_b$, and $\myeta\propto B^2 \propto r^{\mypB}$ at
$r\gg r_b$ with some $\mypB<0$.
A power-law pressure profile $nT\propto r^\mypPP$ would imply
$\mypB=\mypPP$; for an isothermal distribution $\mypB={-2}$.

Such a decline in $\eta$ is found in A2029 above $r_b\lesssim 120\kpc$, and possibly also in the MHs in A1835 and RXJ1347, as seen in
Fig. \ref{fig:MH_RadialTCorrectedProfiles}.
The figure also shows a simple model for A2029, where
we assume that $B^2=knT$ and that $\myeta$ asymptotes to its average GH value at small radii, $\myeta(r\to0)\to \eta_0 = 10^{-4.4}$, so the proportionality constant $k$ is the only free parameter. This fit (dot-dashed curve) corresponds to a central magnetic field $B_0\simeq 2.1\muG$.
However, due to the limited range of data in this cluster, there is a degeneracy between $B_0$ and $\myeta_0$ --- higher magnetic fields are possible if the central value of $\myeta$ in A2029 is lower than the GH average, and vice versa. The exponential fit of M09 (dashed red curve) suggests higher magnetic fields and lower $\eta_0$; a fit with $B_0=6.3\muG$ is shown (dotted) to illustrate this.
The corresponding low value of $\myeta_0$ in this cluster could arise from its low star formation rate, as discussed in \S\ref{sec:CRI_Origin}.

Assuming that $B^2\propto nT$ for $r<r_b$, we can place a
lower limit on the central magnetic field amplitude $B_0$ in each observed halo, given a pressure model.
If no break in $\myeta$ is identified out to the halo radius $R_\nu$, then
\begin{eqnarray} \label{eq:B0}
B_0> B_{0,min} \equiv \left[ \frac{(n T)_{r=0}}{(n T)_{r=R_\nu}} (f_{ic}^{-1}-1) \right]^{1/2} B_{cmb}(z) \coma \,\,\,\,\,
\end{eqnarray}
where we allowed a fraction $f_{ic}\simeq 0.5$ of inverse Compton
losses at $R_\nu$.  The $B_{0,min}$ values of the halos in our sample
are presented in Table \ref{tab:halos}, assuming that $R_\nu=\min\{r_b,3r_e\}$
and adopting individual isothermal $\beta$-models for each cluster, as detailed in the table.
In the MHs, the
central magnetic field could be much higher than $B_{0,min}$
because {\it (i)} the significant growth in $nT$ towards the center
(due to the density cusp) is not captured by the $\beta$-model; and {\it
(ii)} the magnetic field immediately beneath the CFs could be very
strong, near equipartition \citep{KeshetEtAl10}.
However, in MHs with only a declining $\myeta$ profile observed (\eg A2029), adopting the upper limit on $r_b$ may overestimate $B_0$.

Instead or in addition to a radial break in brightness, a halo in which the
magnetic field is marginal, $B\simeq B_{cmb}$, could become clumpy or filamentary, appearing
bright only in $B\gtrsim B_{cmb}$ islands.  This is probably the reason
for the unusual clumpy or filamentary radio morphology observed in
RXC\,J2003.5–-2323 \citep{GiacintucciEtAl09}, A2255, and A2319 (M09).
Indeed, a decline in $\myeta$ is directly seen in A2319 around $r=150\kpc$
(suggesting low magnetization; see Fig. \ref{fig:MH_RadialTCorrectedProfiles}), and
A2255 is the only strongly ($20\%$--$40\%$) polarized GH known
todate \citep{GovoniEtAl05}; the absence of strong beam depolarization
suggests relatively weak magnetic fields \citep[see, for example,][]{MurgiaEtAl04}.
Additional evidence for the low magnetization in these three halos is their relatively steep
spectrum, as discussed in \S\ref{sec:Spectrum}.

Such marginally magnetized halos are particularly interesting because
the magnetic field can be determined in multiple locations, and
because the radio morphology directly gauges the magnetization or
magnetic decay process.
Fast putative variations in the magnetic field may be easier to detect in such regions, as they could involve temporal changes in the small scale radio morphology.
For example, in A2319, SKA could resolve such changes over a light-crossing time of $\sim 2$ years.

While radio emission rapidly declines outside the $B>B_{cmb}$ region,
an opposite signature is expected in inverse-Compton emission from the
same {\CREs}, as they scatter CMB photons to high energies.  This
emission may be clumpy or filamentary in $B\sim B_{cmb}$ regions.  However, as
pointed out by \citet{KushnirWaxman09}, such radiation cannot explain
the hard X-ray excess detected in a number of clusters \citep[for
review, see][]{RephaeliEtAl08}, because the inverse Compton signal
from secondary {\CREs} has $\nu I_\nu$ comparable to that in the radio
(see Eq.~(\ref{eq:local_correlation})), $\sim 3$ orders of magnitude
lower than needed to account for the detected hard X-ray excess.

\subsection{Using the radio spectrum to disentangle {\CRIs} and magnetic fields}
\label{sec:Spectrum}

Spectral steepening with increasing $r$, increasing $\nu$, or
decreasing $T$, has been reported in several radio halos
\citep[][and references therein]{FerettiGiovannini08, FerrariEtAl08, GiovanniniEtAl09}.  Such trends naturally arise in our
model, due to the energy-dependence of the inelastic cross-section for
collisions of {\CRPs} with the intracluster gas at the relevant
$E_p\sim 20\GeV$ {\CRP} energies.

Secondary {\CREs} are mainly produced by charged pion production,
$p+p\to \pi^{\pm}+X$, where $X$ is any combination of particles,
followed by mesonic decays $\pi^\pm\to \mu^\pm+\nu_\mu(\bar{\nu}_\mu)$
and leptonic decays $\mu^\pm\to
e^\pm+\nu_e(\bar{\nu}_e)+\bar{\nu}_\mu(\nu_\mu)$.  Other processes, in
particular ${\rm He}$-$p$ collisions, have a similar cross-section
per nucleon and should not modify our results by more than $10\%$.  On
average, the energies of the {\CRE}, the $\pi^\pm$, and the {\CRP} are related by $E_e
\simeq f_{e|\pi} E_\pi \simeq f_{e|\pi} f_{\pi|p} E_p$, where
$f_{e|\pi}\simeq1/4$ and $f_{\pi|p}\simeq1/5$
\citep{GinzburgSyrovatsky61}.
The inclusive cross section for $\pi^{\pm}$ production varies significantly
as a function of $E_p$ around $E_p\simeq 10\GeV$, dropping from $>40\mb$ above $30\GeV$ to zero at
the threshold energy $0.3\GeV$ \citep{BlattnigEtAl00}.

We compute the radio spectrum based on two different methods. Model A
uses the spectral fits for $e^+$ and $e^-$ production in inelastic
$p$-$p$ scattering according to \citet[][valid for
$0.5\GeV<E_p<512\TeV$]{KamaeEtAl06}, with corrected parameters and
cutoffs (T. Kamae \& H. Lee 2010, private communication). Model B assumes
that $E_e=(E_p/20)$ and utilizes the inclusive cross sections for
charged pion production according to \citet{BlattnigEtAl00}, which
agree with experimental data for $0.3<(E_p/\mbox{GeV})\lesssim 300$
\citep{Norbury09}.  Model A should be more accurate, because
$f_{e|\pi}f_{\pi|p}$ is not independent of $E_e$ and the {\CRP}
spectrum.

\begin{figure}[h]
\centerline{\epsfxsize=8.5cm \epsfbox{\myfig{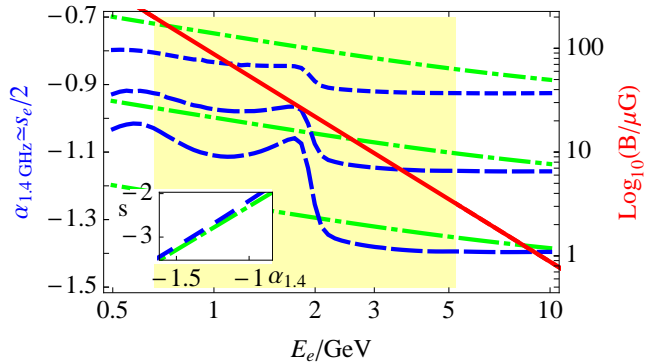}}}
\caption{ The relation between $\alpha$ (dashed curves; left axis),
$B$ (solid curve, right axis) and $E_e$ (abscissa), for $\nu_e=1.4\GHz$.  The
highly magnetized, $B_{cmb}(z=0)<B<200\muG$ regime is shaded.  The radio
spectral index $\alpha_{1.4}$ is shown as a function of $E_e$ according to model A
(dashed) and model B (dash-dotted), for {\CRP} spectral indices $s=-2$,
$-2.5$, and $-3$ (top to bottom).
See \S\ref{sec:Spectrum} for model and parameter description.
\emph{Inset}: The {\CRP} energy index $s$ as a function of radio spectral index $\alpha_{1.4}$ for $B=B_{cmb}$
(for example, at the spatial halo break; see \S\ref{sec:Morphology}).
\label{fig:MHSpectra}
}
\end{figure}

The results, depicted in Fig.~\ref{fig:MHSpectra} for emitted frequency  $\nu_e=(1+z)\nu=1.4\GHz$
and various power-law {\CRP} spectra $n_p(E_p)\propto E_p^{s}$, show
that the radio spectrum steepens with increasing cosmic-ray electron/positron
energy,
\begin{eqnarray} \label{eq:Ee}
E_e\simeq 4.2\nu_{e,1.4}^{1/2}(B/5\muG)^{-1/2}\GeV \fin
\end{eqnarray}
Model A features a spectral break at $E_e\simeq 2\GeV$,
attributed mainly to electron production through diffractive processes, in which one or both protons transitions to an excited
state (discrete resonances and continuum).
Note that even if this channel is blocked, a similar spectral break is found at $E_e\sim 1\GeV$.
A more dramatic steepening than in either model is found if we assume that $E_e=(E_\pi/4)$ and use the pion spectra produced in $p$--$p$ collisions according to \citet{BlattnigEtAl00}; however, these formulae have not been tested beyond $E_p=50\GeV$.

Due to the limited energy range over which the energy index $s_e\equiv d \log(n_e)/d\log(E_e)$ of the injected {\CREs} can be computed, the radio spectrum in Fig. \ref{fig:MHSpectra} is calculated in the approximation where each {\CRE} emits a single photon, such that $\alpha\simeq s_e/2$ \citep[\eg][]{KeshetEtAl03}.
A more accurate computation, convolving the {\CRE} distribution with the synchrotron emission function, would somewhat smear the spectral features.

The radio steepening with increasing $E_e$ implies steepening with
increasing $\nu$ or $r$, or decreasing $T$, as observed.  Both models
agree that spectral steepening by $|\Delta \alpha_\nu|>0.2$
indicates $B>20(\nu_e/1.4\GHz)\muG$ magnetic fields in the region associated with a
flatter radio spectrum.
Measuring $\alpha$ across the $E_e\simeq 2\GeV$ spectral break can be used to
unambiguously fix the {\CRP} spectrum $s$.
Similarly, measuring $\alpha$ at the spatial $\myeta$ break where $B\simeq
B_{cmb}$ can be used to determine $s$, as illustrated in the inset of Fig. \ref{fig:MHSpectra}.
Conversely, if a distinct spectral break exists as predicted by model A, it would directly gauge the magnetic field, once the {\CRP} spectrum has been determined.
Multi-frequency radio data could thus allow a sensitive mapping of the magnetic field, both above and below $B_{cmb}$, throughout the halo, using, for example, deprojected maps at several radio frequencies.

Spectral measurements of radio halos should be interpreted with
caution.  Contamination by {\CRE} sources such as shocks, relics,
central AGNs, and radio galaxies, is common.  Spatial averaging often
blends together different sources, depending on flux sensitivity and
angular resolution.  Therefore, with present observations, the spectrum can be reliably
associated with a halo only when measured locally in uncontaminated regions; in our model these are regions of uniform $\myeta$.
Also note that radio emission above 1 GHz is
increasingly suppressed by the Sunyaev-Zel'dovich effect
\citep{Ensslin02}.

For example, $\myeta(r)$ is nearly constant in A665 within the range $r< r_{max} \simeq 150''$ examined in Fig. \ref{fig:MH_RadialTCorrectedProfiles}. Spectral steepening with increasing $r$ was identified in this cluster \citep{FerettiEtAl04a} by comparing radio maps at $0.3$ and $1.4\GHz$ frequencies. The spectral index steepens from $\alpha\simeq -1.0$ in the halo's center to $\alpha \simeq -1.3$ towards the South and towards the East, within the flat $\myeta$ region. Similar steepening was found towards the North and West beyond $r_{max}$, but the spectrum first flattens to $\alpha\simeq -0.9$.

Similarly, the regular GH in A2744 exhibits a linear radio-X-ray
correlation in brightness out to $r\simeq 1\Mpc$ \citep{GovoniEtAl01}.
Along the main NW elongation, the
$0.3-1.4\GHz$ spectral slope $\alpha(r<200\kpc)\simeq -1.0$,
slightly flattens to $-0.9$ around $500\kpc$, and steepens again to $-1.5$
as $r\to 1\Mpc$. An azimuthal average, however, includes a NE relic tail and
under-threshold regions, leading to an unrealistically uniform
$\alpha(r<\mbox{Mpc})\simeq -1$ \citep{OrruEtAl07}.
A similar behavior is found in A2219 \citep{OrruEtAl07}.

In these examples, the spectral index steepens with increasing $r$ from $\alpha\simeq -1.0$ to $-(1.2$-$1.5)$ in the less perturbed/contaminated direction, sometimes after a mild flattening to $\alpha \simeq -0.9$. Such steepening was also reported in A2163 \citep[]{FerettiEtAl04a} and in Coma \citep{GiovanniniEtAl93}.

Comparable steepening was found as a function of frequency in the GH in A754.
Its spectrum measured between $74$ and $330\MHz$, $\alpha_{0.07}^{0.3}\sim -1.1$, steepens to
$\alpha_{0.3}^{1.4}\sim -1.5$ \citep{BacchiEtAl03}.
More substantial steepening has been reported in other clusters, such as A2319 \citep{FerettiEtAl97} and A3562 \citep{GiacintucciEtAl05}.
However, contamination by extended radio galaxies was reported in these halos.

Significant steepening with increasing $r$ or $\nu$ as in the examples
above is more consistent with the $E_e\sim2\GeV$ spectral break of
model A than with model B.  In model A, strong magnetic fields with $B\gtrsim
10(\nu_e/700\MHz)\muG$ are present in regions where $\alpha_\nu$ is flat.
Moreover, an uncontaminated spectral break measured at some emitted frequency $\nu_b$ would
imply that the local projected magnetic field is $B\simeq 10(\nu_b/700\MHz)\muG$.
Observations at several frequencies can thus be used to map the local projected magnetic field.
As a preliminary example, interpreting the radio spectrum of A754 as a spectral break somewhere between $\nu_e=200$ and $900\MHz$
near the center of the halo, would imply $3<B_0/{\mu\mbox{G}}<14$.

Figure \ref{fig:MHSpectra} indicates that radio steepening from
$\alpha=-1.0$ to $-1.3$ corresponds to a {\CRP} spectral index
$s\simeq -2.7$.  Steepening beyond $-1.5$, if uncontaminated, would imply a steep {\CRP} spectrum with
$s<-3$.  The radio spectrum at high, $E_e\gg 10\GeV$ {\CRE} energies
approaches $\alpha\to s/2$.  However, inward of the spectral break, at
$E_e< 2\GeV$, $\alpha\simeq -1$ depends only weakly on $s$.  This could
explain the universally observed $\alpha\simeq-1$ in the centers of halos.

\begin{figure}[h]
\centerline{\epsfxsize=8.5cm \epsfbox{\myfig{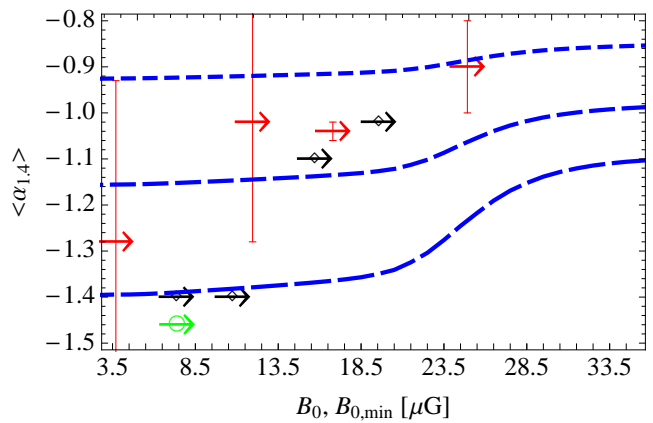}}}
\caption{ The relation between the spatially averaged,
$\nu\lesssim1.4\GHz$ spectral index $\langle \alpha_{1.4} \rangle$, and
the central magnetic field $B_0$.  The measured spectral index in each GH is
plotted as a function of the minimal central magnetic field
$B_{0,min}$ calculated in \S\ref{sec:Morphology} (right arrows; data with no reported error bars shown as diamonds; data with error bars larger than the y-axis shown as a circle; see Table \ref{tab:halos}).
Also shown (dashed curves) is the average spectrum calculated in a simple magnetic equipartition, isothermal
$\beta$-model (see \S\ref{sec:Spectrum}), as a function of the central magnetic field, for {\CRP} energy indices $s=-2$,
-2.5, and -3 (top to bottom).
\label{fig:MHSpectraAve}
}
\end{figure}

Figure \ref{fig:MHSpectraAve} shows the average $\nu\lesssim 1.4\GHz$
spectral indices $\langle \alpha_{1.4} \rangle$ reported for GHs in our
sample, as a function of their minimal central magnetic fields
$B_{0,min}$ calculated in \S\ref{sec:Morphology}.  The results are
also summarized in Table \ref{tab:halos}.  A positive correlation is
seen, in the sense that halos with stronger central magnetization tend to have a flatter
radio spectrum, as expected.  Note that this correlation is stronger than, and
not simply due to, a correlation we find between $\langle \alpha \rangle$ and the halo size.

For comparison, we also plot (dashed lines in Fig. \ref{fig:MHSpectraAve})
the average radio spectrum $\langle \alpha_{1.4}\rangle$ as a function of
$B_0$, computed for a typical isothermal $\beta$-model
with $\beta=0.7$, assuming $B^2\propto nT$, spectral model A, and
radio emission extending to $R_\nu\simeq 3r_c$.
Although this result cannot be quantitatively compared to the GH data
(which are only lower limits on $B_0$), the trends qualitatively agree.
This provides an independent indication that the {\CRP} spectrum is steep,
with $s\lesssim -2.7$.

The radio spectrum is relatively steep in the five GHs and MHs for
which we presented evidence for low magnetization in \S\ref{sec:Morphology}, in agreement with
the $\langle \alpha \rangle-B_0$ correlation.  The halos which are
not shown in Fig. \ref{fig:MHSpectraAve}, include
A2029 \citep[$\langle\alpha_{0.2}\rangle\simeq
-1.35$,][]{SleeSiegman83},
A2255 \citep[$\langle\alpha_{0.3}^{1.4}\rangle=-2.06\pm0.57$,][where
we infer $\langle\alpha_{0.6}^{1.4}\rangle\sim -1.5$ by comparing M09
with \citet{HarrisEtAl80}]{KempnerSarazin01},
Perseus \citep[$\langle\alpha_{0.3}^{1.5}\rangle\sim -1.2$,][]{GittiEtAl04},
and
RXC\,J2003.5–-2323 \citep[$\langle\alpha_{0.2}^{1.4}\rangle\sim
-1.32\pm0.06$,][]{GiacintucciEtAl09}.

Finally, a cautionary comment is in place regarding the extrapolation of the spectrum to low frequencies that correspond to $E_e\lesssim1\GeV$ {\CRE} energies.
Although $E_e=1\GeV$ {\CREs} are produced on average by $E_p\sim 20\GeV$
{\CRPs} (for $s=-2.5$), the contribution of low energy {\CRPs} with $E_p\lesssim 3\GeV$ is not negligible.
We expect a {\CRP} power law energy spectrum $n_p(E_p)\propto E_p^{s}$ to be modified around the proton rest mass.
Consequently, our $s=\constant$ approximation becomes increasingly unlikely at low, $E_e\lesssim \mbox{few}\GeV$ {\CRE} energies.

\section{Discussion: origin of {\CRIs} and magnetic fields}
\label{sec:Discussion}

In the previous sections we studied the luminosity (in \S\ref{sec:Correlation}) and the surface brightness (in \S\ref{sec:RadialProfile}) properties of GHs and MHs, showed that they are best explained by a model invoking secondary {\CREs} and strong magnetic fields as the source of radio halos (in \S\ref{sec:Theory}), and discussed some implications of such a model (in \S\ref{sec:Model}).
Here we consider the origins of the magnetic fields (\S\ref{sec:Magnetization}) and of the primary {\CRIs} (\S\ref{sec:CRI_Origin}). We subsequently summarize our work in \S\ref{sec:Summary}.

\subsection{Magnetization: mergers (GHs) and sloshing (MHs)}
\label{sec:Magnetization}

In the model derived above, the primary {\CRI} population is
long-lived and has very similar properties in different clusters and in different parts of a cluster.
Hence, the defining property of halos is sufficiently high
magnetization.  Measuring strong, $B>B_{cmb}$ magnetic fields without radio
detection at the level given by Eq.~(\ref{eq:local_correlation}) would
rule out the model, unless the {\CRI} energy fraction is exceptionally low (\eg due to a low level of star formation; see \S\ref{sec:CRI_Origin}).
The model can be tested by checking if the different magnetic field estimates it produces in a given halo are self consistent, and agree with independent
measurements.

In the model, GH clusters are associated with strong, $B\gtrsim B_{cmb}\sim 3\muG$ magnetic fields, whereas weaker, $B<B_{cmb}$ fields are present in clusters devoid of a halo.
Magnetic fields of the order of $B_{cmb}$, extending over Mpc scales, are consistent with Faraday rotation measures (RM) in non-cool clusters, considering the uncertainties involved.
Such RM studies suggest magnetic fields ranging between a few $\mu$G to $10\muG$, extending out to $r\sim 500\kpc$, in a sample of non-cool core clusters, on $10$--$20\kpc$ coherence scales \citep{Clarke04}.
Weaker magnetic fields, typically $\sim 1\muG$ to a few $\mu$G, were found using other methods, such as simulating the polarization of extended sources (Murgia et al. 2004) or using a Bayesian maximum likelihood analysis of the Faraday RM \citep{VogtEnsslin05}.

RM studies could potentially test the association between strong magnetization and the presence of halos, as implied by the model, although this is complicated by several factors.
First, RM analyses involve substantial systematic and statistical errors associated with projection effects, the magnetic power spectrum, the range of coherence scales, and non-Gaussianity.
Second, cluster fields are often reconstructed using peripheral RM sources, assuming some magnetic field scaling $B\propto n^\mu$ that extends both inside and outside the halo; however this assumption is uncertain and in some cases inconsistent with our model, \eg in asymmetric or clumpy halos.
Finally, the RM gauges the magnetic field amplitude integrated along the line of sight, whereas the radio emission scales differently with the magnetic field, \eg as $B^2$ in $B\lesssim B_{cmb}$ regions.

It is interesting to compare the magnetic fields estimated in different clusters. Care must be taken to compare the same measures estimated under the same assumptions.
For example, assuming a Gaussian, Kolmogorov magnetic power spectrum with $\mu=1/2$, we find $\ave{B(r<3r_c)}=1.8\muG$ and $B(3r_c)=1.3\muG$ in Coma \mbox{\citep{BonafedeEtAl10}},  which harbors a GH, while weaker fields, $\ave{B(r<3r_c)}=1.2\muG$ and $B(3r_c)=0.8\muG$ are found in A2382 \mbox{\citep{GuidettiEtAl08}}, which has no halo. Similarly, $\ave{B(r<3r_c)}=1.3\muG$ and $B(3r_c)=0.8\muG$ are found in A119 \mbox{\citep{MurgiaEtAl04}}, which does not harbor a GH, but this result is obtained for $\mu=0.9$ so comparison to the other two clusters may be misleading. Note that it is not clear, considering the inherent uncertainties, if the differences between these $B$ estimates are significant.

The association between GHs and merger events is fairly well
established \citep[see][and references therein]{FerettiGiovannini08}.
With present magnetic field estimates, it
is quite plausible that a major merger event could amplify the
intracluster magnetic field by a factor of a few, sufficient to exceed $B_{cmb}$ and produce a GH.  The model thus
resolves the notorious discrepancy between magnetic field estimates
based on Faraday RM and on previous GH analyses, which unnecessarily
assumed equipartition between magnetic fields and {\CREs}.

The very strong magnetic fields we infer in the centers of MHs are
consistent with the RM observed in CCs, which typically imply $B\simeq
10-40\muG$ in the core, on coherence scales of a few up to $10\kpc$
\citep{CarilliTaylor02}.  It is interesting to point out that the
cooling flow power is correlated both with the RM \citep{TaylorEtAl02}
and with the MH radio power \citep{GittiEtAl04}, while no strong correlation
between MH and AGN power was identified \citep[\eg][]{GovoniEtAl09}.

The association between MH edges and CFs strongly suggests that
sloshing motions play a major role in magnetizing the core.
Notice that $\sim 10\kpc$ scales are indeed
characteristic of the shear magnetic amplification anticipated beneath
CFs \citep{KeshetEtAl10}.  Sloshing could suppress cooling in the core,
for example by mixing the cold gas with a heat inflow
\citep{MarkevitchVikhlinin07}.  In such a scenario, a stronger cooling
flow may correspond to a larger magnetized region, leading to correlations
between the cooling flow power and both RM and MH power, as observed, but not
necessarily to an MH-AGN correlation.

The distribution of MHs (without the AGN component) among CCs should reflect the statistics of core magnetization, and therefore of sloshing.
We expect a correlation between the presence of MHs and of CFs in a cluster, and a correlation between the MH size or power and the shear flow strength, manifest for example in the number and size of CFs and in the \citep[measurable, see][]{KeshetEtAl10} shear across them.
Assuming that some level of magnetization by core sloshing is always present in CCs, as suggested by the ubiquity of observed CFs, the steep magnetic profile associated with the density cusp would imply that every CC has some MH, even if small.  The MH distribution would then be continuous, and not bimodal as in GHs.

Shear amplification of magnetic fields parallel to the CF plane is expected mostly below the CF and in a thin boundary layer around it, in which the field can reach high, near equipartition levels \citep{KeshetEtAl10}.
The morphological association between MH edges and CFs, discovered by \citet{MazzottaGiacintucci08}, is therefore expected to be ubiquitous.
It is best seen where CFs are observed edge on; elsewhere it may be observed as morphological correlations between radio maps and spatial X-ray gradients, which trace projected CFs.

The magnetic field (the polarization) is expected to be parallel (perpendicular) to the CF, \ie approximately tangential (radial).
Polarization would be preferentially observable where beam deprojection is minimal, \ie at large radii where the magnetic field weakens.
Interestingly, nearly radial, $10\%-20\%$ polarization was detected in the MH in A2390, growing stronger with increasing radius \citep[we refer to the spherical, $r\sim 150\kpc$ component around the cD galaxy in this irregular MH; see][]{BacchiEtAl03}.
We predict that near CF edges, where the magnetic field is particularly strong, the radio spectrum would be relatively flat (see \S\ref{sec:Spectrum}).

\subsection{{\CRI} Origin: diffusion and supernovae sources}
\label{sec:CRI_Origin}

The spectral steepening of the radio signal with increasing {\CRE} energy $E_e\propto(\nu/B)^{1/2}$, provides a novel method for measuring
the primary {\CRP} spectrum.
The radio steepening observed in some halos, roughly from $\alpha \simeq -1.0$ to $\alpha\simeq -1.3$, corresponds to a {\CRP} spectral index $s\lesssim -2.7$ at $E_p\sim20\GeV$ energies.
The {\CRP} energy fraction then becomes (\cf Eq.~\ref{eq:injection})
\begin{eqnarray} \label{eq:uCR}
\xi_{p}(>E_p) & \equiv & \frac{u_{p}(>E_p)}{u_{th}}
\simeq 10^{-3.6\pm0.2}\left(\frac{E_p}{10\GeV}\right)^{-0.7} \fin \,\,\,\,
\end{eqnarray}
Note that with the uncertain and possibly contaminated radio spectra presently available, a steeper {\CRP} spectrum with
$s\simeq -3$ is possible.  The model would be challenged if the uncontaminated spectrum of a substantial halo population turns out to be much steeper than $\alpha=-1.5$, unless the corresponding steep {\CRP} spectrum can be explained.

The {\CRP} distribution in Eq.~(\ref{eq:uCR}) resembles (but has an energy fraction a few $100$ times smaller than) the {\CRP} distribution found in the solar vicinity above $1\GeV/$nucleon.
This distribution could originate from sources that
inject roughly equal energy per decade of {\CRP} energy ($s_0\simeq
-2.2$), such as supernovae
(SNe), if energy-dependent diffusion is significant in the inner halo
regions.  For example, a simple estimate of {\CRI} scattering off
magnetic irregularities with a Kolmogorov power spectrum yields a
diffusion coefficient $D\simeq
10^{30}(E_p/\mbox{GeV})^{1/3}(B/\mu\mbox{G})^{-1/3}\cm^2\se^{-1}$
\citep{VolkEtAl96}.  This implies {\CRI} diffusion over $\sim 0.5\Mpc$ during
a Hubble time, and a steepening by $\Delta s=-1/2$.
More substantial steepening is possible if the diffusion function has a stronger energy dependence. For example, the diffusion function is often assumed to scale as $D\propto E_p^{1/2}$, which could lead to a $\Delta s=-3/4$ steepening in the {\CRI} spectrum.

The {\CRI} output of SNe can be crudely estimated \citep{VolkEtAl96}
if we assume that a fraction $f_{\mbox{\scriptsize{II}}}$ of the
cluster's $Z=0.3Z_{0.3}$ solar metallicity is seeded by Type II SNe,
which on average produce $0.1M_\odot M_{Fe,0.1}$ of iron and deposit a
fraction $\xi=0.3\xi_{0.3}$ of the $10^{51}E_{51}\erg$ explosion
energy in $E_p>10\GeV$ {\CRIs},
\begin{eqnarray} \label{Eq:SNeCRI}
\xi_{p}^{SN} \simeq 0.03 f_{\mbox{\scriptsize{II}}} Z_{0.3} E_{51} M_{Fe,0.1}^{-1} \xi_{0.3} \fin
\end{eqnarray}
This can reproduce Eq.~(\ref{eq:uCR}) if over the cluster's lifetime,
the {\CRIs} diffuse to distances a few times larger than the radius $R_\nu$ of
the radio halo.
Note that if {\CRI} diffusion is entirely absent, the {\CRIs} accelerated in SNe
would be confined to the cluster, and adiabatic losses could only lower their
energy density to the level of Eq.~(\ref{eq:uCR}).
However, they would then retain their flat, $s\simeq -2.2$ spectrum.

An SNe origin of {\CRIs} can be tested by examining the correlations
between $\myeta$ and (intensive) tracers of SNe activity among different halos.
One possible tracer is the local metallicity measured at $r=0.1r_{500}$, tabulated in Table \ref{tab:halos}.
We chose to use $Z(0.1r_{500})$ because \emph{(i)} it was measured for all the M09 halos with \emph{XMM-Newton} profiles in \citet{SnowdenEtAl08}; \emph{(ii)} the spatially averaged $Z$ is not meaningful when temperature gradients are large; and \emph{(iii)} $0.1r_{500}$ lies well within the $Z(r>0.02r_{500})\propto r^{-0.3}$ decline typically found in both cool and non-cool core clusters \citep{SandersonEtAl09}.
While our sample is statistically small, Perseus, which shows a significantly higher $\myeta$ than in all the other halos in our sample, also shows a slightly higher $Z(0.1r_{500})$ \citep{SnowdenEtAl08}.
However, due to the large uncertainty in abundance measurements, the elevated metallicity in Perseus is not significant ($<1\sigma$) with respect to some halos.
Moreover, at smaller radii $r\lesssim 0.025r_{500}$, the metallicity in A2029 appears to be higher than in all other halos, and is significantly ($>3\sigma$) higher than in Perseus \citep{SnowdenEtAl08}.

Although better metallicity statistics may identify a more significant correlation between $\myeta$ and $Z$, metallicity is probably not the most useful tracer of the SNe contribution to {\CRIs} in the halo.
Metallicity provides a cumulative measure of SNe activity, tracing the metals released from all past SNe in the cluster. In a model where a significant fraction of the {\CRIs} have already diffused away from the cluster's center, metallicity would not linearly trace the population of {\CRIs} residing within the halo, especially in the more compact MHs.
It is more appropriate to use an intensive tracer of \emph{recent} SNe activity, such as the star formation rate (SFR) normalized by the cluster's gas mass $M_g$, or the fraction of star forming galaxies. A correlation between an SNe measure and a {\CRI} tracer, such as $\myeta$ or the deviation from the luminosity correlation $\nu P_\nu/\bar{L}_{X[0.1,2.4]}^{1.7}$, may be more useful than the metallicity in establishing or ruling out an SNe origin of halo {\CRIs}.

As seen in Fig.~\ref{fig:MH_RadialTCorrectedProfiles},
the halos with the highest $\myeta$ in our sample are the MH in Perseus and the GH in A665, while the halos with the lowest $\myeta$ are the MH in A2029 and the GH in A773.
Interestingly, the literature shows evidence for exceptionally high specific star formation in both Perseus and A665, and for a low specific SFR in A2029.
(We found no relevant data for A773.)

In Perseus, which has the smallest $M_g$ \citep[by at least a factor of $4$, see][]{FukazawaEtAl04} and one of the most powerful cooling flows within our sample \citep[\eg][]{White00,AllenEtAl02}, there is optical-to-UV evidence for a relatively high SFR \citep[\eg][]{BregmanEtAl06, RaffertyEtAl08}.
In particular, the cD galaxy NGC1275 in Perseus has a high SFR of $\sim 30M_\odot \yr^{-1}$ \citep{DixonEtAl96} --- the highest in our sample when normalized by $M_g$.
Notice that the central galaxy in A1835 has a higher SFR of $\sim 100M_\odot \yr^{-1}$ --- the highest SFR known in such objects \citep{PetersonFabian06}.
However, $M_g$ is $\sim 10$ times larger in A1835 with respect to Perseus \citep{FukazawaEtAl04}.
Note that regions containing a high density of cosmic-rays are directly observed in Perseus in the form of X-ray cavities, reaching distances $r>100\kpc$ (M. Markevitch, private communications).
A high specific SFR is also inferred in A665, which was found to be the cluster with the highest dispersion in color magnitude relation (a known tracer of star formation) in a sample of 57 X-ray bright clusters \citep{LopezCruzEtAl04}.
In contrast, A2029 has a low specific SFR, as it was shown to have a SFR $\sim 70$ times lower than in A1835 \citep{HicksMushotzky05} while its gas mass is only $\sim 1.8$ times smaller \citep{FukazawaEtAl04}.

While these trends support an SNe origin of {\CRIs}, more work is needed in order to quantify their significance and compile a comparative statistical analysis.
Note that the combination of a strong correlation of $\myeta$ with the specific SFR and a poor correlation with metallicity, if established, would directly imply that {\CRI} diffusion is significant.
Indeed, it is difficult to explain the steep, $s\lesssim -2.7$ spectrum without invoking {\CRI} diffusion.

The above estimates of diffusive steepening assume that most of the {\CRIs} produced by the sources presently dominating the halo have already escaped beyond it. This is consistent with the typical SFR peak at $z\sim 1$, and with the above estimates of the halo {\CRI} abundance and the total {\CRI} output of SNe (\cf Eqs.~\ref{eq:uCR} and \ref{Eq:SNeCRI}).
However, such substantial diffusion would introduce some scatter in the radio--X-ray correlation, depending on the {\CRI} production history of each cluster. Quantitative estimates of the SNe history of GH clusters, needed to compute this scatter, are beyond the scope of this work.

\subsection{Summary and Conclusions}
\label{sec:Summary}

We have shown that the radio-X-ray correlation in GH luminosity
(Eq.~\ref{eq:GH_Correlation}) can be generalized
(Eq.~\ref{eq:GeneralizedLx}) to hold for both GHs and MHs
(Fig.~\ref{fig:MH_CorrectedCorrelation}), by correcting for the
halo size.  A universal, linear relation between the radio and X-ray surface brightness, $\eta=10^{-4.4\pm0.2}$, was presented (Eq.~\ref{eq:local_correlation} and Fig.~\ref{fig:MH_RadialTCorrectedProfiles}).
This, combined with the radial $\myeta$ and $T$ profiles, implies a universal radio emissivity $\nu j_\nu = 10^{-31.4\pm0.2} n_{-2}^2 (T/T_0)^{0.2\pm0.5} \erg\se^{-1}\cm^{-3}$ (Eq.~\ref{eq:nu_j_nu}) near the center of halos.  We argued that these results and their applicability to GHs and MHs alike, strongly support one model for all halos, involving secondary {\CREs} (injected according to Eq.~\ref{eq:injection}) and strong magnetic fields with $B\gtrsim B_{cmb}$, while disfavoring other models (Table \ref{tab:models}).

This model makes useful predictions without requiring additional
assumptions or fine tuning.  Radio emission rapidly fades in regions where $B$ drops below $B_{cmb}$, producing a distinct radial break (\eg in A2029 and in Perseus; Fig \ref{fig:MH_RadialTCorrectedProfiles}) or a
clumpy/filamentary radio morphology (\eg in RXC\,J2003.5–-2323, A2255,
and A2319) that can be used to map $B\simeq B_{cmb}$ contours.
Marginally magnetized regions with $B\lesssim B_{cmb}\propto (1+z)^2$ are characterized by relatively high polarization and a steeper radio spectrum; their morphology traces the magnetic evolution and can potentially reveal a temporal signal.
We expect a higher incidence rate of such transition regions at higher redshift, while no halos should exist at very high redshift.

Another direct consequence of the model is radio spectral steepening with increasing $E_e^2\propto \nu/B$ (Eq.~\ref{eq:Ee}), \ie with increasing $r$ or $\nu$ or decreasing T, as indeed is observed.  Such steepening, and in particular a $B\simeq 10(\nu_e/700\MHz)\muG$ spectral break
(Fig. \ref{fig:MHSpectra}), gauges the magnetic field,
roughly producing an additional $B$ contour for each radio map frequency.
The spectral break could be used to accurately map $B$ throughout the halo, using future radio telescopes such as the Murchison Widefield Array (MWA\footnote{http://www.mwatelescope.org}), the LOw Frequency ARray (LOFAR\footnote{http://www.lofar.org}), and SKA.

A pressure model can be used to extrapolate $B$ throughout
the cluster. This indicates central magnetic fields $B_0$
(Eq.~\ref{eq:B0}) that exceed $10\muG$ in most halos (see Table
\ref{tab:halos} for lower limits $B_{0,min}$).  A correlation between
the average radio spectral index $\langle \alpha\rangle$ and $B_0$, implied by the model, was identified in GH data (Fig.~\ref{fig:MHSpectraAve}).

In our model, any source of strong ($B\gtrsim B_{cmb}$), persistent magnetic fields in the intracluster medium
would have similar properties to radio halos, as long as it does not significantly inject additional cosmic rays.
This may include some extended radio galaxies, which were recently found to exhibit properties similar to halos \citep{RudnickLemmerman09}.
Conversely, the universal value of $\myeta$ we predict for any highly magnetized, uncontaminated region in the intracluster medium provides a powerful test of the model.

The spectral steepening of the radio signal, the universality of $\alpha\simeq-1$ in the
centers of halos, and the correlation between $\langle \alpha \rangle$
and $B_{0,min}$ (Fig. \ref{fig:MHSpectraAve}), indicate a steep {\CRI}
spectrum, $s\lesssim-2.7$, and thus favor significant {\CRI} diffusion.
In a diffusion model, the most plausible source of the {\CRIs} (Eq.~\ref{eq:uCR}) is SNe (\eg Eq.~\ref{Eq:SNeCRI}).
We show (in \S\ref{sec:CRI_Origin}) preliminary evidence for a correlation between $\myeta$ and the SFR normalized by the gas mass $M_g$, supporting an SNe {\CRI} origin.
None of these properties is expected in an alternative model (K09), in which the secondary {\CREs} arise from $\sim 1\GeV$ {\CRPs}, which are accelerated in the cluster's virial shock and advected inward with the flow, thus being compressed to $\sim 10\GeV$ energies.
Note that the data slightly favors a $j_\nu\propto n^2T^0$ scaling within each cluster (see \S\ref{sec:Theory}), which is natural if {\CRIs} originate in SNe, rather than the $j_\nu\propto n^2T^1$ behavior anticipated if they are accelerated in the virial shock.
Also note that adiabatic compression of {\CRIs} produced at the virial shock and advected with the gas would lead to a radially increasing, $\myeta\propto n^{-1/3}$ profile (due the soft equation of state of relativistic particles), which is not observed (see Fig.~\ref{fig:MH_RadialTCorrectedProfiles}).

We stress that although our model and the model of K09 disagree regarding the origin of {\CRIs}, the {\CRI} distribution, the spectral properties of halos, and the role of diffusion, we reach the same conclusions regarding the radio mechanism:  emission from secondary {\CREs} in strong magnetic fields.
This conclusion is based on
\emph{(i)} the tight radio-X-ray correlation in total GH luminosity and the GH bimodality (K09);
\emph{(ii)} the tight radio-X-ray correlation in both coincident luminosity and surface brightness, in both GHs and MHs, despite their different physical properties;
\emph{(iii)} the strong magnetic fields inferred from Faraday RMs in MHs and possibly (see \S\ref{sec:Magnetization}) also in GHs;
\emph{(iv)} the tightening of the brightness correlation at small radii, away from merger shocks, radio relics, and their associated turbulence;
\emph{(v)} the $j_\nu\propto n^2 T^\mypT$ scaling of the radio emissivity within each halo, where $\mypT\lesssim 1$;
\emph{(vi)} the coincidence between MH edges and CFs, manifest as a sharp radial cutoff in $\myeta$ (\eg in Perseus);
\emph{(vii)} a power-law radial break where $\myeta(r\ll r_b)\simeq\constant$ and $\eta(r\gg r_b)\propto B^2$, possibly seen in the MHs in A2029, A1835 and RXJ1347, and in the GH in A2319;
\emph{(viii)} the clumpy/filamentary morphology of some halos, where independent evidence for low magnetization is present;
\emph{(ix)} the spectral steepening and the correlation between $\langle \alpha \rangle$ and $B_0$ (this suggests strong magnetization provided that the {\CRI} spectrum is steep, $s\lesssim -2.7$).

Each aspect of our model can be tested in the near future.
The association between the presence of halos and strong, $B\gtrsim B_{cmb}$ magnetic fields can be directly tested by comparing the magnetization levels independently estimated in halo and in non-halo clusters, as illustrated in \S\ref{sec:Magnetization}.
The secondary origin of the halos can be tested if the {\CRIs} are detected directly through their $\pi^0$ production; for example, such a detection of a {\CRI} component substantially stronger than in Eq.~(\ref{eq:uCR}) would rule out our model.
The SNe origin of the {\CRIs} can be tested by carefully examining the correlation between a {\CRI} measure such as $\eta$, and an intensive SNe tracer such as the specific SFR, in a sample of halo clusters.

\acknowledgements
We are deeply grateful to Maxim Markevitch for many
fruitful discussions.
We thank Ramesh Narayan, Doron Kushnir, Boaz Katz, Eli Waxman, Yuying Zhang, Annalisa Bonafede, Gianfranco Brunetti, Julius Donnert, and Matteo Murgia for useful comments.
UK acknowledges support by NASA through
Einstein Postdoctoral Fellowship grant number PF8-90059 awarded by the
Chandra X-ray Center, which is operated by the Smithsonian
Astrophysical Observatory for NASA under contract NAS8-03060. This
work was supported in part by NSF grants AST-0907890, AST-08 and NASA
LUNAR grant for AL.

%\bibliography{\mybib{Cluster}}
% Abbrev M09 and K09.

\end{document}